\documentclass[11pt]{article}
\usepackage[a4paper,top=2.5cm,bottom=2.5cm,left=3cm,right=3cm]{geometry}
\usepackage[a4paper]{geometry}
\usepackage[T1]{fontenc}
\usepackage[utf8]{inputenc}                 
\usepackage{microtype}                     
\usepackage[english]{babel}
\usepackage{emptypage}
\usepackage{indentfirst}
\usepackage{booktabs}  
\usepackage{tabularx} 
\usepackage{listings} 
\usepackage[font=small]{quoting}  
\usepackage{amsmath,amssymb,amsthm}                                   
\usepackage{chngpage,calc}   
\usepackage[dvipsnames]{xcolor}    
\usepackage{lipsum}
\usepackage{eurosym} 
\usepackage{hyperref} 
\usepackage{indentfirst}
\usepackage{authblk}

\usepackage{natbib}

\usepackage[noend]{algpseudocode}
\usepackage{dsfont}
\usepackage{float}
\usepackage[caption = false]{subfig}
\usepackage[final]{graphicx}
\usepackage{tasks}
\usepackage{courier}
\usepackage{gensymb}
\usepackage{multirow}
\usepackage[shortlabels]{enumitem}
\usepackage{gensymb}
\usepackage{bm}

\usepackage{marginnote}
\reversemarginpar

\usepackage{graphicx}

\DeclareMathOperator*{\argmin}{arg\,min}

\makeatletter
\def\BState{\State\hskip-\ALG@thistlm}
\makeatother

\usepackage{titlesec}
\usepackage[dvipsnames]{xcolor}
\usepackage{epstopdf}
\usepackage{wrapfig}
\usepackage{verbatim}
\usepackage{pbox}
\usepackage{fancyhdr}
\usepackage{float}
\usepackage{adjustbox}
\usepackage{tikz} 

\makeatletter
\newcommand*{\centerfloat}{%
	\parindent \z@
	\leftskip \z@ \@plus 1fil \@minus \textwidth
	\rightskip\leftskip
	\parfillskip \z@skip}
\makeatother
\usepackage{acro}
\usepackage{setspace}

\acsetup{first-style=short}
\usepackage{ragged2e}

\usepackage{minipage-marginpar}
\usepackage{placeins}
\usepackage{float}

\usepackage{listings}
\usepackage{afterpage}

\usepackage{algorithm2e}

\usepackage[flushleft]{threeparttable}

\usepackage{graphicx}

\usepackage{mathrsfs,upref}

\theoremstyle{definition}

\theoremstyle{remark}

\usepackage{enumitem}
\newlist{steps}{enumerate}{1}
\setlist[steps, 1]{label = \textbf{Step \arabic*}:}

\makeatletter
\newcommand*{\rom}[1]{\expandafter\@slowromancap\romannumeral #1@}
\makeatother
\numberwithin{equation}{section}

\title{A bias-adjusted estimator in quantile regression for clustered
  data}
 \author[1]{Maria Laura Battagliola}
 \author[1,*]{Helle S\o rensen}
 \author[1]{Anders Tolver}
 \author[2]{Ana-Maria Staicu}
\affil[1]{Department of Mathematical Sciences, University of Copenhagen\\ \smallskip}
\affil[2]{Department of Statistics, North Carolina State University\\ \smallskip}
\affil[*]{Corresponding author: helle@math.ku.dk}
\date{March 26, 2021}

\begin{document}
\maketitle
\begin{abstract}
\noindent
The manuscript discusses how to incorporate random effects for
quantile regression models for clustered data with focus on settings
with many but small clusters. The paper has three contributions: (i)
documenting that existing methods may lead to severely biased
estimators for fixed effects parameters; (ii) proposing a new two-step
estimation methodology where predictions of the random effects are
first computed {by a pseudo likelihood approach (the LQMM method)} and
then used as offsets in standard quantile regression; (iii) proposing
a novel bootstrap sampling procedure in order to reduce bias of the
two-step estimator and compute confidence intervals. The proposed
estimation and associated inference is assessed numerically through
rigorous simulation studies and applied to an AIDS Clinical Trial
Group (ACTG) study.  \color{black} \medskip

\end{abstract}


\section{Introduction}
\noindent
Quantile regression has been introduced by \cite{koenker1978regression}
as a way to describe the association between covariates and quantiles of the response distribution at pre-set quantile levels. See the comprehensive monographs by \cite{koenker2005} and \cite{koenker2017handbook} on quantile regression. \textcolor{black}{In recent years, quantile regression has for example been employed in econometrics and finance \citep{BAYER201856,WANG2018115,MACIAK2021,MACIAK2021_1}.} 
In this article we consider linear quantile regression for clustered data, such as longitudinal data, and discuss estimation approaches that properly account for the inherent dependence of the observations within the same cluster. Research in this area has been very active, especially in econometrics, but existing methods for quantile regression estimation are proved to be asymptotically consistent only when both the number of clusters and cluster size increase to infinity.  This assumption is rather strong in practice, where the common scenario is that there are many clusters of  moderate to small sizes. When the cluster size is small, numerical investigations show (see Figure \ref{fig:plot1}) that the popular quantile regression estimators may exhibit severe bias, even if there are many clusters. This represents a  gap in the literature, as data settings that involve many clusters of small to moderate sizes are ubiquituos in medicine and animal science, to name a few.

Existing approaches to account for dependence in parameter estimation of quantile regression for clustered (repeated measures) data treat the cluster-specific parameters either as fixed or random. 
For example, \cite{RePEc:eee:econom:v:170:y:2012:i:1:p:76-91} and
 \cite{GALVAO201692} use cluster-specific intercepts 
and estimate them as fixed effects parameters together with the quantile regression parameters using the so-called fixed effects quantile regression (FE-QR) and fixed effects smoothed quantile regression (FE-SQR), respectively, while \cite{GALVAO20151} and \cite{10.1093/jjfinec/nbx016} develop minimum-distance-based estimation for the same purpose. Some approaches consider shrinkage to deal with an increasing number of clusters, in the presence of cluster-specific parameters.  Penalized quantile regression for longitudinal data 
is discussed by \cite{koenker2004quantile}, \cite {lamarche2010robust}, \cite{doi:10.1002/jae.2520} and \cite{GU2019}.
\cite{doi:10.1111/j.1368-423X.2011.00349.x} proposes a two-step estimator, relying on mean regression estimates of cluster-specific intercepts, see also \cite{RePEc:cfr:cefirw:w0249}.
\citet{geraci2007quantile} and \citet{geraci2014linear} introduce a pseudo likelihood approach, where a linear quantile mixed model (LQMM) with random cluster parameters is used as a working model, and \citet{galarza2017quantile} develop an EM-based estimation methodology for the LQMM  framework.
\citet{doi:10.1198/073500107000000269} discuss estimation in a model with correlated random effects (CRE), and \cite{Luo2012} consider a fully Bayesian quantile inference using Markov Chain Monte Carlo, to account for correlated random effects.  
We consider a frequentist perspective and propose a novel two-step estimation approach and associated inference that rely on the LQMM framework.

When the cluster-specific parameters are treated and estimated as fixed effects parameters, estimation suffers from what is known in the literature as the ``incidental parameters problem'' \citep{neyman1948consistent, Lancaster2000}: the number of (nuisance) parameters grows with the number of clusters, leading to inconsistent joint estimation, when the cluster size is small. Not surprisingly, only asymptotic scenarios where both the number of clusters and the cluster size increase to infinity have been studied \citep{koenker2004quantile, RePEc:eee:econom:v:170:y:2012:i:1:p:76-91, GALVAO201692, doi:10.1111/j.1368-423X.2011.00349.x, RePEc:cfr:cefirw:w0249}. 
To bypass the issues caused by the incidental parameter problem, the cluster-specific parameters can be modeled as random effects; however, asymptotic properties are not studied for the LQMM-based estimator \citep{geraci2007quantile, geraci2014linear}.

Different solutions have been suggested for bias-adjustment in the case of small clusters: 
\cite{GALVAO201692} introduce an analytical adjustment for FE-SQR based on asymptotic analysis, nonetheless the approach requires an optimal bandwidth selection, which is challenging in practice.  The authors also adapt the half-panel jackknife method \citep{dhaene2015split} to longitudinal quantile regression. We consider the use of half-panel jackknife for bias correction in our numerical investigation.
%
Usually, bootstrap methods have been used for construction of confidence intervals in models with cluster-specific effects \citep{galvao2015bootstrap, doi:10.1111/j.1368-423X.2011.00349.x, geraci2014linear},
and for marginal models (without cluster-specific effects), see for example
\citet{Karlsson2009} and \cite{doi:10.1080/01621459.2016.1148610}. 
We introduce a non-standard bootstrap technique for both bias-adjustment and inference of quantile regression parameters, in the context of clustered (longitudinal) data. 

This paper makes three main contributions. First, we numerically demonstrate that Koenker's penalized estimator, Canay's two-step estimator and the LQMM estimator can be severely biased when clusters are small or of moderate size. Although no papers have claimed the opposite, we are the first to raise this issue.
Second, we propose a new estimation methodology and associated inference for the quantile regression parameters.
The point estimator is computed in two steps: (i) an LQMM framework is used to predict the cluster-specific parameters; and (ii) the predictions are used as offsets in a standard quantile regression. The two-step estimator is furthermore adjusted for bias using bootstrap, and the third contribution is the novel combination of wild bootstrap and ordinary resampling, that reduces bias and allows to construct confidence intervals that have good coverage performance. Numerical studies show that the proposed estimator has considerably smaller bias than the existing competitors, when the cluster size is small.  


The structure of the paper is as follows: we set up the model framework in Section~\ref{sec:model}. In Section~\ref{sec:estimation} we  summarize some of the existing estimation methods in quantile regression for repeated measures data and then present the proposed estimation method. The estimation method is evaluated numerically in a thorough simulation study in  Section~\ref{sec:simulation} (with additional results in the appendix) and applied to a clinical trial regarding HIV treatments in Section~\ref{sec:application}.
The paper concludes with Section~\ref{sec:conclusion}, which discusses the main findings.

\section{Regression framework}\label{sec:model}

\noindent
Let $(Y_{ij}, x_{ij})_{j=1}^{n_i} $ be the observed data for the $i$th cluster 
($i=1,\ldots,N$), where $x_{ij}\in \mathbb{R}^{p-1}$ is the vector of covariates corresponding to the $j$th observation of the $i$th cluster and $Y_{ij}\in \mathbb{R}$ is the respective response. Here $n_i$ denotes the cluster size and the responses are assumed independent across different clusters but expected to be correlated within the same cluster. Let $\tau \in (0,1)$ be a fixed quantile level of interest, and let $Q^i_{Y_{ij}|x_{ij}}(\tau)$ be the $\tau$th quantile of the conditional distribution of $Y_{ij}$ given $x_{ij}$ for cluster $i$. Consider a linear quantile regression model

\begin{align}
\label{eqn:linear_quantile}
Q^i_{Y_{ij}|X_{ij}}(\tau) = X_{ij}^T \beta_i^\tau,
\end{align}
where $X^T_{ij} = (1, x_{ij}^T)$ and
{$\beta_i^\tau=(\beta_i^{\tau,1}, \ldots, \beta_i^{\tau,p})$} is an unknown vector regression parameter that
quantifies the association between the covariates and the
$\tau$-quantile of the response for cluster $i$. Due to the definition of $X_{ij}^T$, the first component of $\beta_i^\tau$ is the intercept; by an abuse of notation we refer to $X_{ij}$ as the vector of covariates.

This model formulation allows for cluster-level effects for every scalar component of $X_{ij}$; an equivalent formulation is to represent the cluster-level effect as the sum of a population level effect and a cluster-specific deviation. Such formulation is standard in the mixed effects model representation \citep{laird1982random}, and we adopt it here as well.
As for mean regression, all covariates are not necessarily modeled with cluster-specific levels, and the selection of variables without cluster-specific effects can be based on interpretational as well as computational arguments. 
Without loss of generality, assume that only the first $q\leq p$ components of $X_{ij}$ have cluster-varying effects; denote by $Z_{ij}$ the vector formed by the first $q$ elements of $X_{ij}$.  The remaining $n-q$ components of $X_{ij}$ have only population level effect. The effects corresponding to $Z_{ij}$ are used to account for the dependence of the observations within the same cluster; for example, \cite{koenker2004quantile}, \cite{doi:10.1111/j.1368-423X.2011.00349.x}, and \cite{galvao2017quantile} used a random intercept only ($q=1$) to model this dependence.  Using the terminology from linear mixed effects we can re-write model (\ref{eqn:linear_quantile})
as
\begin{equation} \label{eq:qregfull}
	Q^i_{Y_{ij}|X_{ij}}(\tau) = X_{ij}^T \beta^\tau + Z_{ij}^T u_i^\tau,
\end{equation}
by separating the quantile regression parameters that describe a population level effect, $\beta^\tau=(\beta^{\tau,1},\ldots, \beta^{\tau,p})$, from the ones that describe cluster-specific deviations, $u_i^\tau=(u_i^{\tau,1},\ldots, u_i^{\tau,q})$. Just like in linear mixed models, it is assumed that $u_i^\tau$ are zero mean random quantities. Our primary interest lies in the estimation of $\beta^\tau$ in situations with many clusters (large $N$) but modest cluster sizes (small $n_i$s).

\color{black}


Let $\mathbf{u}^\tau = (u^\tau_1, \ldots, u_N^\tau)$ denote the collection of (unobserved) cluster-specific parameters.
Moreover, let $\mathbf{Y}$ be the vector of the (observed) responses $Y_{ij}$.
Consider the loss function
\begin{equation}\label{eq:L}
L(\beta^\tau, \mathbf{u}^\tau; \mathbf{Y}) = \sum_{i=1}^N \sum_{j=1}^{n_i} \rho_\tau(Y_{ij}- X_{ij}^T\beta^\tau -Z_{ij}^Tu_i^\tau ),
\end{equation}
where $\rho_\tau(v)=v(\tau - \mathbf{1}_{(v<0)})$ is the check function \citep{koenker1978regression}. 
If the values of the cluster-specific effects, $u_i^\tau$, were observed,
a natural estimator would be the linear quantile regression estimator corresponding to the covariates $X_{ij}$ and the modified responses $Y_{ij}-Z_{ij}^Tu^\tau_{i}$.
We call this the oracle estimator,
\begin{align}\label{eq:oracle}
\hat \beta_{\text{oracle}}^{\tau} = \underset{\beta^{\tau}}{\argmin} \hspace{1mm}L(\beta^\tau,\mathbf{u}^\tau; \mathbf{Y});
\end{align}
evidently the estimator $\hat \beta_{\text{oracle}}^{\tau}$ enjoys the asymptotic properties of a standard quantile regression estimator \citep{koenker2005}. However, $\hat \beta_{\text{oracle}}^{\tau}$ is an
unattainable estimator, as $u_i^\tau$s are not observed, and the question we consider in this paper concerns the effect of uncertainty in the cluster-specific effects on estimating the population level quantile regression parameter.

One way to address the estimation problem is to treat $u_i^\tau$s in \eqref{eq:qregfull} as fixed effects parameters and have them estimated jointly with $\beta^\tau$ using a standard quantile regression framework. The FE-QR estimation of \citet{RePEc:eee:econom:v:170:y:2012:i:1:p:76-91} minimizes the  loss function \eqref{eq:L} with respect to both $\beta^\tau$ and $\mathbf{u}^\tau$. With this approach, the number of parameters grows at the same rate as the number of clusters, so the estimator of $\beta^\tau$ is only consistent in asymptotic scenarios where $n_i$ grows faster than $N$ \citep{RePEc:eee:econom:v:170:y:2012:i:1:p:76-91}. 

We will instead pursue an approach to estimate $\beta^\tau$, when $u_i^\tau$s are treated as random. 
Similar to the generalized linear mixed effects framework, there are two interpretations of the covariates' effects on the response distribution quantile. On one hand, we have the conditional perspective, following from the definition \eqref{eq:qregfull} that
$P( Y_{ij} \leq X_{ij}^T\beta^\tau + Z_{ij}^T u_i^\tau|X_{ij},u_i^\tau)=\tau$, which states that $\beta^\tau$ is the quantile regression parameter associated with the covariates $X_{ij}$, conditional on the cluster-specific effects. On the other hand, we have the marginal perspective that $P( Y_{ij} \leq X_{ij}^T\tilde \beta^\tau|X_{ij})=\tau$, which describes the covariates' effect on the $\tau$-quantile of the marginal distribution of $Y_{ij}$. The two quantile regression parameters ($\beta^\tau$ and $\tilde{\beta}^\tau$) are generally different in the same manner that a fixed effects parameter of a generalized linear mixed model has a different interpretation than its counterpart in a marginal or population average approach  \citep{zeger1988, neuhasu1991}. The difference between the conditional and marginal quantile models is discussed more thoroughly in \citet{reich2009}, see also the simulation model in Section~\ref{sec:simulation}.

As a consequence, also pointed out in \citet{koenker2004quantile}, it is vital for the estimation of $\beta^\tau$ of a conditional perspective that the cluster-specific parameters $u_i^\tau$ are not ignored. Indeed, we illustrate in Section \ref{sec:compare} that the
simple marginal quantile regression estimator
$
\hat{\beta}^\tau_{\text{marg}} = \arg \min_{\beta^\tau} L(\beta^\tau, \mathbf{0}; \mathbf{Y}) = \arg \min_{\beta^\tau} \sum_{i,j} \rho_\tau(Y_{ij}-X_{ij}^T \beta^\tau) 
$
based on standard quantile regression (where all $u_i$s are replaced by zero) may be severely biased for $\beta^\tau$.

The conditional perspective implies that
$$
P ( Y_{ij} - Z_{ij}^T u_i^\tau \leq X_{ij}^T\beta^\tau |X_{ij}) = \tau,
$$
where the probability is taken with respect to the joint distribution of $Y_{ij}$ and $u_i$. Inspired by this equality, we propose to first predict the cluster-specific effects and then use these predictions as offset in a standard linear quantile regression model using a transformed response. 

\section{Estimation}\label{sec:estimation}

\subsection{Review of selected methods for estimation and
  bias-adjustment}\label{sec:existingmethods}

\subsubsection*{Penalization of cluster-specific parameters}

\noindent
The model \eqref{eq:qregfull} was first introduced in the literature by \citet{koenker2004quantile} in a simpler form, where the term $Z_{ij}^T u_i^\tau$ is replaced by only a cluster-specific intercept, call it $u_{i0}$, which is assumed to be quantile-invariant. For fixed  quantile level $\tau$, both the parameter $\beta^\tau$ and the cluster-specific intercepts, $u_{i0}$, are estimated by minimizing the penalized loss function
\begin{equation}
  \label{pen_crit}
  L(\beta^{\tau}, \mathbf{u}_0; \mathbf{Y})+ \lambda \sum_{i=1}^N |u_{i0}|,
\end{equation}
where $\lambda \geq 0$ is a regularization parameter. \citet{koenker2004quantile} uses $\ell_1$ penalty in \eqref{pen_crit} due to its computational convenience;
in our numerical investigation of the estimators in Section~\ref{sec:simulation}, we also use  $\ell_2$
penalty and find minor differences.  While (\ref{pen_crit}) focuses on a single quantile level, \citet{koenker2004quantile} describes the estimation of the quantile regression parameters  simultaneously at multiple quantile levels, by introducing quantile-level weights and minimizing a weighted penalized likelihood.

The $\ell_1$-penalized estimator for $\beta^{\tau}$ is consistent and
asymptotically normal, provided that $N^a/n\to 0$ for some $a>0$
(where $n_i=n$); see \citet{koenker2004quantile}. Nonetheless, when
the cluster size, $n_i$, is small the estimator may not enjoy these
theoretical properties and can be seriously biased, especially for
extreme quantile levels; see Section~\ref{sec:simulation}.

\subsubsection*{Canay's two-step estimator}
\noindent
\citet{doi:10.1111/j.1368-423X.2011.00349.x} assumes a cluster-specific intercept, $u_{i0}$, too, but considers a two-step
procedure to estimate the linear quantile regression parameter
$\beta^\tau$ of \eqref{eq:qregfull}. First, $u_{i0}$ are estimated as
part of the fixed parameters in a mean regression framework. Second,
the quantile regression parameter $\beta^\tau$ is estimated using a
standard quantile regression framework \citep{koenker1978regression}
applied to adjusted responses  $\widetilde Y_{ij} = Y_{ij} - \hat
u_{i0}$, where $ \hat u_{i0}$ denotes the estimated cluster-specific
effects from the previous step. Equivalently, $\beta^\tau$ is estimated by minimizing the loss function \eqref{eq:L} with $Z_{0i}=1$ and $\mathbf{u}^\tau$ replaced by $\hat
{\mathbf{u}}_{0}$, the vector containing the $\hat u_{i0}$s:
$$
    \hat\beta^\tau_{\text{Canay}} = \argmin_{\beta^\tau} L(\beta^\tau,
    \hat{\mathbf{u}}_{0}; \mathbf{Y}).
    $$
\cite{doi:10.1111/j.1368-423X.2011.00349.x} and
\cite{RePEc:cfr:cefirw:w0249} discuss asymptotic properties for
$\hat\beta^\tau_{\text{Canay}}$ in scenarios where both the number of clusters and cluster
size increase.

The use of the mean regression in the first step is justified in Canay's
set-up because only intercepts are allowed to be cluster-specific, and
the deviations from the average are assumed to be constant over
quantile levels.
In such case, the random effects correspond to location shifts; their estimation is quantile-invariant, which may be restrictive. 
Moreover, while treating $u_{i0}$s as fixed parameters as opposed to random may lead to negligible differences, in terms of estimation, for large clusters, 
the correct approach for small clusters is to treat them as random parameters. To address this issue, 
we propose a new quantile regression estimator in Section~\ref{sec:twostep-boot}, which is inspired by \citet{doi:10.1111/j.1368-423X.2011.00349.x}.

\subsubsection*{Marginalization over random effects in a working model (LQMM)}

\noindent
\citet{geraci2007quantile, geraci2014linear} propose to embed the problem in a 
fully specified working model, a linear quantile mixed model (LQMM), using the duality between the quantile loss (check function) and the asymmetric Laplace distribution (ALD, 
\cite{doi:10.1080/03610920500199018}). Specifically, assume $u_i \sim f(\cdot; \varphi)$ for some density $f$ that is parameterized by a scale parameter $\varphi$ and
posit the following joint model for the responses $Y_{ij}$s and the cluster-specific $u_i$s:
\begin{equation}
\label{LQMM_model}
\begin{gathered}
Y_{ij} | u_i, X_{ij} \stackrel{ind}{\sim} ALD(X_{ij}^T\beta^{\tau}+Z_{ij}^Tu_i, \sigma,
\tau), \hspace{5mm} j=1,...,n_i\\
u_i \stackrel{iid}{\sim}  f(\cdot,\varphi),
\end{gathered} 
\end{equation}
for $i=1,...,N$ , where $\sigma$ is a scale parameter for the residual
distribution. The conditional $\tau$-quantile function associated to
the working model is given by (\ref{eq:qregfull}), and the conditional
likelihood of $Y_{ij}$s given $X_{ij}$s and $u_i$s takes the form
\eqref{eq:L}; with $u_i^\tau=u_i$.

Estimation of model parameters $(\beta^\tau, \sigma, \varphi)$ is based on maximizing the pseudo likelihood of $\textbf{Y}$ obtained by integrating the joint density of $(Y_{i1},\ldots,Y_{in_i}, u_i)$ with respect to the distribution of latent random effects $u_i $.
In practice, the
random effects are assumed to be drawn either from a Gaussian
distribution $N(0, \varphi^2)$ or a Laplace distribution $ALD(0,
\varphi,1/2)$, see \citet{JSSv057i13} for details about
the computations. In the special case of random intercepts only, 
when the Laplace distribution is used for the cluster-specific parameters $u_{i0}$s, 
maximizing the joint model (\ref{LQMM_model}) is equivalent to minimizing Koenker's penalized loss function, 
while if the Gaussian distribution is used, then maximizing the joint model (\ref{LQMM_model}) is equivalent to minimizing the $\ell_2$-penalized criterion. From this perspective, the tuning parameters using Koenker's penalization approach are scale parameters in the joint model framework and thus can be estimated with increased computational efficiency. 
Finally, once the parameters $\beta^\tau$, $\sigma$ and $\varphi$ are
  estimated, the random effects
  can be predicted using best linear predictors (BLPs),
see equation (12) in \cite{geraci2014linear}.
These predictions are essential ingredients
for the new estimator suggested in
Section~\ref{sec:twostep-boot}; note that the computed predictions
vary with the level $\tau$ even though $u_i$ in the model
\eqref{LQMM_model} does not. 



\cite{geraci2007quantile} and \cite{geraci2014linear} do not discuss asymptotics for the LQMM
estimator, but if the working assumptions are true (ALD for the within-cluster distribution and Gaussian
or Laplace distribution for the random effects), then the
LQMM estimator is the maximum likelihood estimator, and the usual
asymptotic results hold.
On the other hand, the bias of the LQMM
estimator may be non-negligible, even when $N$ is large, 
if the data generating process does not coincide with the working
model. This will be illustrated in Section~\ref{sec:compare}.

\subsubsection*{Jackknife-based bias-adjustment for an existing estimator}
\noindent
Since the estimators above show bias when used for clustered data, a
bias reduction adjustment would be appropriate. 
There are various ways to do this; one approach to reduce the bias of
an estimator is by using a jackknife bias-adjustment. 
 The half-panel jackknife was first introduced in
 \citet{dhaene2015split} as a method for bias correction for mean
 regression in longitudinal settings with many subjects and fixed
 panel size. 
 Later, it was {applied to the FE-SQR
   estimator for longitudinal quantile
 regression \citep{GALVAO201692}}; we describe it here for clustered data.

We randomly split the dataset into two sub-datasets, each containing half of the observations from every cluster. Denote the quantile regression estimator from the two sub-datatsets by $\hat{\beta}^{\tau}_{1}$ and $\hat{\beta}^{\tau}_{2}$, respectively, and let $\hat{\beta}^{\tau}$ be
the estimator from the full dataset.
Then, the half-panel jackknife estimator $\hat{\beta}^{\tau}_{\text{jackknife}}$ is
defined as
\begin{equation}
\label{jack}
\hat{\beta}^{\tau}_{\text{jackknife}}
= \hat{\beta}^{\tau} - \left( \frac 12(\hat{\beta}^{\tau}_{1} + \hat{\beta}^{\tau}_{2})-\hat\beta^\tau\right)=
2\hat{\beta}^{\tau} - \frac{(\hat{\beta}^{\tau}_{1} + \hat{\beta}^{\tau}_{2})}{2}.
\end{equation}
To gain some intuition about the bias reduction  of this estimator, assume that all clusters have equal size $n$ and that the
asymptotic bias of the initial estimator $\hat\beta^\tau$ is of the form
$C/n+o(n^{-1})$ for some constant $C$. Then the asymptotic bias of the jackknife estimator $\hat{\beta}^{\tau}_{\text{jackknife}}$
is of order $o(n^{-1})$, so the order of the bias is reduced. Nonetheless, empirical studies indicate that while the adjustment indeed reduces the bias, the resulting variance of the estimator is increased; see \citet{GALVAO201692}.

\subsection{Proposed quantile estimation with reduced bias}\label{sec:twostep-boot}

\subsubsection*{A new two-step estimator (unadjusted)}

\noindent
We propose to estimate the linear quantile regression parameter
$\beta^\tau$ using a new approach, which is inspired by the LQMM estimation framework and \citet{doi:10.1111/j.1368-423X.2011.00349.x}. It consists of two steps:

\begin{enumerate}[align=left]
	\item[\textbf{Step 1}:] Use the LQMM framework to predict the cluster-specific random effects by the best linear
          predictors (BLPs) and center them; denote the centered prediction for cluster $i$ by $\tilde u_i^\tau$;
	\item[\textbf{Step 2}:] Transform the responses to $\widetilde{Y}_{ij} =Y_{ij} - Z_{ij}^T \tilde u_i^\tau$ and use the standard quantile regression framework for the new responses $\widetilde{Y}_{ij}$ and covariates $X_{ij}$ to estimate $\beta^\tau$.
\end{enumerate}


There are two key differences between the proposed approach and
\citet{doi:10.1111/j.1368-423X.2011.00349.x}: 1) Canay estimates the cluster-specific effects using a mean
regression framework, whereas we use a quantile regression model, and 2) Canay estimates the cluster-specific effects by treating them as fixed parameters; in contrast we view and estimate them as random parameters. We illustrate in Section~\ref{sec:simulation} that these differences have a large impact in terms of the estimation quality of quantile regression parameters.

\begin{figure}[htb]
	\centering
             	\includegraphics[width=9cm]{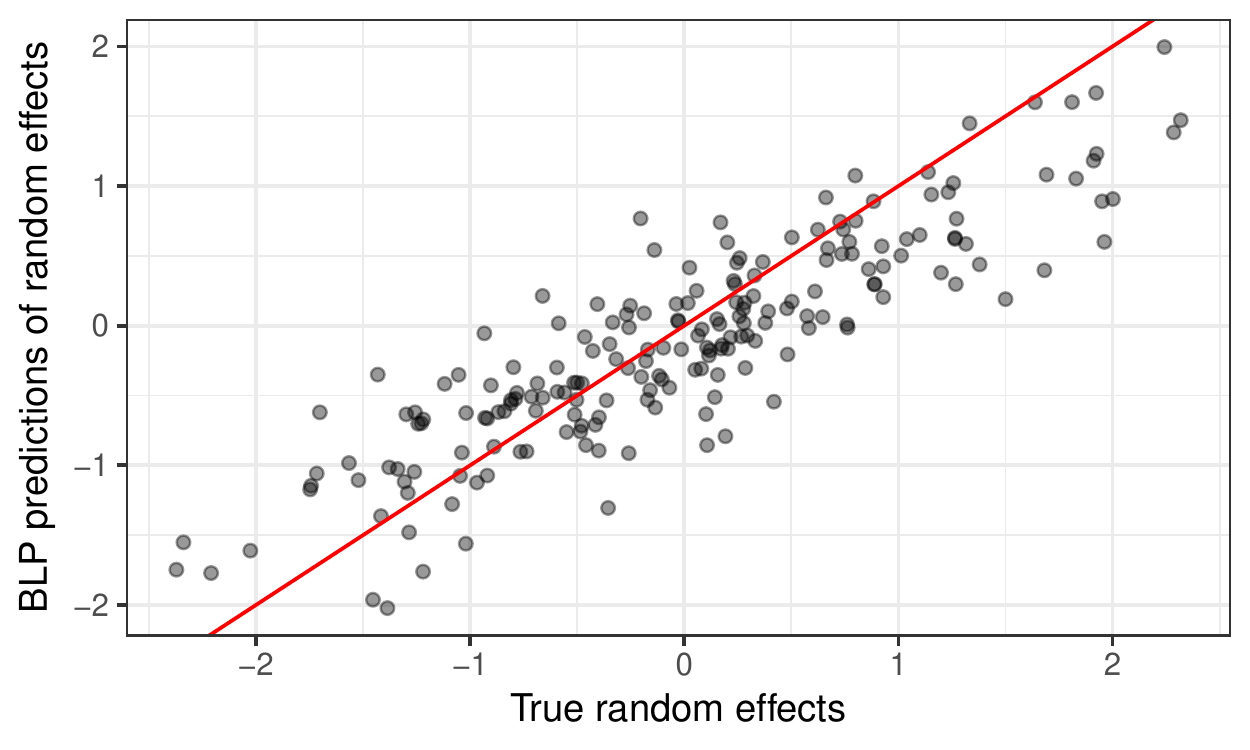} 		
        	\caption{Comparison of the true random effects (on
                  $x$-axis) and centered
		BLP predictions (on $y$-axis) for the first 
		cluster from 200 simulated datasets from the standard
		scenario in Section~\ref{sec:simulation}. The red line is the line with slope one through the origin.}
	\label{fig:uest}
\end{figure}

Figure \ref{fig:uest} shows a comparison between true random
effects ($x$-axis) and their predicted values ($y$-axis) for the first cluster
from 200 simulated data sets representing the benchmark scenario in
Section~\ref{sec:simulation}. The BLPs capture the variation among clusters quite well, but it is
clear that some degree of shrinkage takes place as more extreme random effects are
drawn towards zero.

The second step consists of standard quantile regression
applied to $Y_{ij} - Z_{ij}^T \tilde u_i^\tau$; equivalently the quantile regression parameter is estimated by minimizing the loss function \eqref{eq:L}, with ${\mathbf{u}}$ fixed at value $\tilde{\mathbf{u}}^{\tau}$, the vector containing $\tilde{u}_i^\tau$s:
\begin{align*}
\hat{\beta}^{\tau}_{\text{two-step}} = \underset{\beta^{\tau}}{\argmin} \hspace{1mm}L(\beta^{\tau}, \tilde{\mathbf{u}}^{\tau}; \mathbf{Y}).
\end{align*}

Our two-step estimator turns out to have considerably smaller bias than the LQMM
estimator; yet, the deviation between the true and estimated random
effects introduces some bias. To bypass this issue, we propose a
bias-corrected adjustment based on bootstrap as explained below. 
The second step can be carried out with standard
software, which typically provides standard
errors 
for each component of the vector $\beta^\tau$. However, it is important to
recognize that these uncertainty estimates are not necessarily reliable, as they
only account for the sampling variability of $\hat\beta^\tau _{\text{two-step}}$
conditional on the random effects, not for the extra variation due to
the uncertainty in predicting the random effects. We propose to use
bootstrap to estimate the total variation of $\hat\beta^\tau
_{\text{two-step}}$. We describe the bootstrap procedures used for
bias-adjustment and estimation of variability in the following.

\subsubsection*{Bootstrap sampling for bias-adjustment}


\noindent We propose a semi-parametric-type of bootstrap, which
combines non-parametric bootstrap and wild bootstrap and relies on the
linearity of the quantile regression model. Let $\mathcal{U} =
\{\tilde{u}_1^{\tau}, \ldots, \tilde{u}_N^{\tau} \}$ be the sample of
predicted cluster-specific effects obtained with two-step estimation procedure and for each $i$ and $j$ denote the observed residuals by $\varepsilon_{ij} = Y_{ij} -
X^T_{ij}\hat \beta^\tau_{\text{two-step}} - Z^T_{ij}\tilde{u}_i^{\tau}$. 

We define the bootstrap sample as $ \{  (Y^*_{ij}, X_{ij}, Z_{ij} )_{j=1}^{n_i}, u_i^{\tau,*} \}_{i=1}^N$  
where $u_i^{\tau,*}$s are obtained by resampling with replacement from $\mathcal{U}$ and $Y^*_{ij}$ is defined by
\begin{equation}\label{eq:Ystar}
	Y_{ij}^* = X^T_{ij}\hat \beta_{\text{two-step}}^{\tau} + Z^T_{ij}u_i^{\tau,*} + \varepsilon^*_{ij}, \hspace{5mm} i=1,...,N, \hspace{2mm} j=1,...,n_i,
\end{equation}
where $\varepsilon^*_{ij}$s are attained by wild
bootstrap; see \citet{wu1986} and \citet{liu1988} who introduced this method in the context of mean
regression. Specifically, 
let $\varepsilon^*_{ij}=w_{ij}|\varepsilon_{ij}|$, where $w_{ij}$s are
drawn independently from the following distribution: 
\begin{equation}
	\label{eq:w_distr}
	w=\left\{
	\begin{array}{@{}ll@{}}
		2(1-\tau), & \text{with probability}\ 1-\tau \\
		-2\tau, & \text{with probability}\ \tau
	\end{array}\right. 
\end{equation}
which has the $\tau$-quantile equal to $0$. The idea of scaling the residuals by weights drawn from an asymmetric distribution was proposed by \citet{10.1093/biomet/asr052}; as \cite{wang2018} also
recognized, the wild bootstrap 
captures asymmetry and homoscedasticity better than ordinary
resampling of residuals. Notice that the coupling
  between covariates and residuals is maintained in the equation
\eqref{eq:Ystar}  in the sense that
  each residual is used to generate a bootstrap value for its own observation.


  
Bootstrap methods have been used for inference on quantile regression for longitudinal data. 
Most of the approaches rely on non-parametric resampling where complete
clusters are sampled with replacement, by sampling the covariates and the outcomes jointly 
 \citep{doi:10.1111/j.1368-423X.2011.00349.x,
  RePEc:eee:econom:v:170:y:2012:i:1:p:76-91, galvao2015bootstrap,
  geraci2014linear, Karlsson2009}.
This method is useful for evaluation of an estimator's variation, and thus for 
computation of standard errors and confidence intervals. However, we 
expect such bootstrap estimators to be centered around the estimate
from the observed data, and they would therefore not be useful for bias-adjustment. 
In contrast, our bootstrap procedure ensures that the resampled observations are generated from a distribution with $\hat
\beta_{\text{two-step}}^{\tau}$ as the ``true'' parameter; therefore, we can measure bias as the deviation
between $\hat \beta_{\text{two-step}}^{\tau} $ and the bootstrap estimates. Details
are given below. Our proposed bootstrap method (abbreviated RW, for standard Resampling and Wild) is compared with resampling of complete clusters and two additional approaches in Section~\ref{sec:simulation}.

The RW bootstrap sampling procedure ensures that, conditional on the
resampled random effects, the model assumption about the association between the covariates and
the quantile at level $\tau$ is satisfied with
$\beta^\tau=\hat\beta^\tau_{\text{two-step}}$ (obtained from the
observed data).
Furthermore, if the random effects were known then
all observations were independent, and the distribution of the
bootstrap estimators obtained with wild bootstrap would represent the
sampling distribution of $\hat\beta^\tau_{\text{two-step}}$
\citep{10.1093/biomet/asr052, wang2018}.
However, due to the potential deviation between the working model in
LQMM and the true data generating model, the empirical distribution of LQMM predictors of the random
effects may not fully represent the cluster-to-cluster variation, and
since this variation is driving the bias, the proposed
estimator does not completely remove the bias of the initial estimator asymptotically.

Once a bootstrap sample is available, the quantile regression
estimator is obtained by using the proposed two-step estimation
approach. At this part, information about the resampled
cluster-specific effects are ignored; nonetheless these terms are used in a subsequent step, when we estimate the estimator's variability. The bootstrap estimate of the quantile regression parameter is obtained by averaging the estimates in $B$ such bootstrap samples.  If $\hat\beta^{\tau,*}_{\text{two-step},b}$ denotes the $b$th bootstrap replicate then the overall bootstrap estimate of the quantile regression parameter is
$\bar{\beta}^{\tau, *}_{\text{two-step}} = \sum_{b=1}^B \hat
\beta^{\tau,*}_{\text{two-step},b}/B$. 
The deviation $\bar{\beta}^{\tau,
  *}_{\text{two-step}}-\hat\beta^\tau_{\text{two-step}}$ between the
overall bootstrap estimate and the original estimate is regarded
as an estimate of the bias, so an adjusted estimator  \cite[Chapter
10.6]{EfronTibshirani1993} is defined by
\begin{align} \label{eq:biascorrection}
  \hat\beta^{\tau}_{\text{adj}} 
  =   \hat\beta^\tau_{\text{two-step}} -
  \left( \bar{\beta}^{\tau,
  *}_{\text{two-step}}-\hat\beta^\tau_{\text{two-step}}\right)
  =
	2{\hat\beta^\tau_{\text{two-step}}} -
	\bar{\beta}^{\tau, *}_{\text{two-step}}.
\end{align}
As illustrated by numerical studies, this quantile regression estimator has reduced bias compared to the (unadjusted) two-step estimator. 


\subsubsection*{Confidence intervals}

\noindent
An important advantage of using a bootstrap-based estimator is that it
allows to study the variability of the estimator, and we now discuss construction of the confidence intervals for the quantile regression parameter for each component $k$ of the $p$-dimensional parameter $\beta^\tau$. We consider two approaches: the first approach is based on the so-called basic bootstrap method to construct confidence intervals and the second approach capitalizes on the availability of the bootstrap sample of the cluster-specific effects, which is obtained at each step of the bootstrap procedure.

The basic bootstrap $100(1-\alpha)\%$ confidence intervals \cite[eq. 5.6]{DavisonHinkley1997}
for $\beta^\tau_k$ 
are defined as 
\begin{equation*}
\label{CI_basic}
\left(2{\hat\beta^\tau_{\text{two-step},k}}-\beta _{1-\alpha
    /2,k}^{\tau,*}\, ; \, 2{\hat\beta^\tau_{\text{two-step},k}}-\beta
  _{\alpha /2,k}^{\tau,*}\right), \quad k=1,\ldots,p,
\end{equation*}
where $\beta _{\alpha/2,k}^{\tau,*}$ and $\beta _{1-\alpha/2,k}^{\tau,*}$ are the $\alpha/2$ and $(1-\alpha/2)$ 
quantiles, respectively, in the bootstrap sample of $\hat{\beta}^{\tau,*}_{\text{two-step},k}$.

The second approach to construct confidence intervals relies on a
normal asymptotic distribution for the quantile regression estimator
and the bootstrap-based estimate of the variance of the quantile
regression estimator. However, in contrast to most bootstrap-based
confidence intervals constructed this way, the bootstrap standard
error alone, $\text{SD}_{\text{two-step},k} =\sqrt{\sum_{b=1}^B (
  \hat \beta^{\tau,*}_{\text{two-step},k,b}-  \bar{\beta}^{\tau,
    *}_{\text{two-step},k} )^2   /(B-1)} $, fails to accurately
quantify the full variability of the quantile regression estimator of
$\beta^\tau$. This is due to the shrinkage phenomenon of the LQMM
predicted cluster-specific effects, which is further perpetuated in
the bootstrap samples of $u_i^{\tau, *}$s and incorporated in the
bootstrap replicates $\hat \beta^{\tau, *}_{\text{two-step},b} $.

To bypass this issue, we consider an adjustment. In this regard,
denote by $\text{SE}_{\text{obs},k}$ the estimated standard error of
the $k$th component of $\hat\beta^\tau_{\text{two-step}}$ reported by
the standard quantile regression \citep{koenker1978regression} with
the cluster-specific effects set to the LQMM predicted values and
using the accordingly transformed data 
(step 2 of our procedure).
Recall that this quantity ignores the variability of the
cluster-specific effects, and thus underestimates the true variability
of the regression estimator. Fortunately, our bootstrap algorithm, by
resampling from the empirical distribution of the predicted
cluster-effects, allows us to track the variability of the regression
estimator induced by the uncertainty in predicting these effects. Let
$\hat\beta^{\tau,*}_{\text{oracle},b}$ denote the oracle-type quantile
regression estimator based on the $b$th bootstrap sample, i.e.\ the
$Y^{*b}_{ij}$s, and by using the  ``true'' values of the
cluster-specific effects, i.e.\ the $u_i^{\tau,*b}$s.  
As before, for each component $k$ denote by $\bar{\beta}^{\tau, *}_{\text{oracle}} = \sum_{b=1}^B \hat \beta^{\tau,*}_{\text{oracle},b}/B$ and $\text{SD}_{\text{oracle},k} =\sqrt{\sum_{b=1}^B (  \hat \beta^{\tau,*}_{\text{oracle},k,b}-  \bar{\beta}^{\tau, *}_{\text{oracle},k} )^2   /(B-1)} $ the mean and standard deviation, respectively, of the oracle-type quantile regression estimator.

We define the adjusted standard error of the $k$th component of the two-step quantile regression estimator as
$$
   \text{SE}_{\text{adj},k} = 
   \text{SD}_{\text{two-step},k} \  \frac{\text{SE}_{\text{obs},k}}{\text{SD}_{\text{oracle},k}}
   \quad
   k=1,\ldots,p.
   $$
Since both terms of the ratio are based on keeping the
cluster-specific constant, the ratio is used to account for the
shrinkage phenomenon. Another way to understand the adjusted standard
error is to view it as a multiplicative factor to the standard error
that is reported in our step 2, $\text{SE}_{\text{obs},k}$: in this
case the ratio $
\text{SD}_{\text{two-step},k}/\text{SD}_{\text{oracle},k}$ measures
the extra variation of the quantile regression estimator due to
estimation of the random cluster-specific effects. 
\color{black}

The $100(1-\alpha)\%$ confidence intervals for $\beta^\tau_k$ based on the adjusted standard errors are computed as
  \begin{equation}
  \label{CI_adj}
       \hat\beta^{\tau}_{\text{adj},k} \pm q_{1-\alpha/2}  \cdot \text{SE}_{\text{adj},k},
   \end{equation}
where $q_{1-\alpha/2}$ is the $(1-\alpha/2)$ quantile of $N(0,1)$. These
confidence intervals will later be referred to as SE-adjustment
confidence intervals.

\medskip

We summarize our procedures for estimation and inference in Algorithm~\ref{alg:pseudocode}. 
\begin{algorithm}[h!]
	\label{alg:pseudocode}
	
	Consider data $\{(Y_{ij}, {X}_{ij}, {Z}_{ij})_{j=1}^{n_i}:i=1, \ldots, N\} $\;
	Using LQMM framework, obtain the centered BLPs of the random effects: $\{ \tilde{u_i}^\tau \}$\;
	Use data $\{(\widetilde {Y}_{ij}  , {X}_{ij})_{j=1}^{n_i}:i=1, \ldots, N \}$, where $ \widetilde Y_{ij}   = Y_{ij}-{Z}_{ij}^T\tilde{u_i}^\tau$
        and get the (unadjusted) estimate,
        $\hat{\beta}^\tau_{\text{two-step}}$, and its estimated standard error,
        $\text{SE}_{\text{obs}}$\;
       For all $i,j$ compute residuals as $\varepsilon_{ij} = Y_{ij} - X^T_{ij}\hat \beta^\tau_{\text{two-step}} -Z^T_{ij}\tilde{u}_i^{\tau}$\;
	\ForAll {$b = 1:B$}{
                Draw weights $w_{ij}$ from the weight distribution \eqref{eq:w_distr}\; 
		Use wild bootstrap on $\varepsilon_{ij}$: 
		$\varepsilon_{ij}^{*b} = w_{ij} |\varepsilon_{ij}|$ \;
		Resample $\tilde{u_i^\tau}$ with replacement to get $u_i^{\tau,*b}$\;
		Construct the bootstrap sample: $[\{ (Y_{ij}^{*b}, X_{ij}, Z_{ij})_{j=1}^{n_i},  u_i^{\tau,*b}\}: i=1, \ldots, N]$ where $Y_{ij}^{*b} =
		{Z}_{ij}^T u_i^{\tau,*b} +
		{X}_{ij}^T\hat{\beta}^\tau_{\text{two-step}} +
		\varepsilon_{ij}^{*b}$\;
		Use data 
		$\{(\widetilde {Y}^{*b}_{ij}  , {X}_{ij})_{j=1}^{n_i}:i=1, \ldots, N \}$, where $ \widetilde Y_{ij}^{*b}   = Y_{ij}^{*b}-{Z}_{ij}^Tu_i^{\tau,*b}$
		and standard linear quantile regression estimation to get  $\hat{\beta}_{\text{oracle},b}^{\tau,*}$\;
		Use data 
		$\{( {Y}^{*b}_{ij}  , {X}_{ij}, Z_{ij})_{j=1}^{n_i}:i=1, \ldots, N \}$ and the proposed two-step estimation to get 
		$\hat{\beta}_{\text{two-step},b}^{\tau,*}$\;
	}
        	Compute the two-step bootstrap mean,
                $\bar{\beta}^{\tau, *}_{\text{two-step}} $\;
                For each component $k=1, \ldots, p$, calculate the standard deviation for the two-step and oracle estimators, $ \text{SD}_{\text{two-step},k}$ and $\text{SD}_{\text{oracle},k}$, respectively;
        	
        	For specified $\alpha$, for each component $k=1, \ldots, p$ in part calculate:
        	\begin{itemize}
        		\item[--] $100(1-\alpha)\%$ basic confidence interval:	$\left(2{\hat\beta^\tau_{\text{two-step},k}}-\beta _{1-\alpha
        			/2,k}^{\tau,*}\, ; \, 2{\hat\beta^\tau_{\text{two-step},k}}-\beta
                              _{\alpha /2,k}^{\tau,*}\right)$
                            \vspace*{-0.3cm}
        		\item[--]  $100(1-\alpha)\%$ SE adjusted confidence interval: $\hat\beta^{\tau}_{\text{adj},k} \pm q_{1-\alpha/2}  \cdot \text{SE}_{\text{adj},k}$, where $ \text{SE}_{\text{adj},k} = 
        		\text{SD}_{\text{two-step},k} \  \frac{\text{SE}_{\text{obs},k}}{\text{SD}_{\text{oracle},k}}$
        	\end{itemize}
	\bigskip
	\caption{Pseudo code for implementation of the bootstrap
		adjusted two-step estimator and related confidence intervals.}
\end{algorithm}

\subsection{Software}\label{sec:software}

\noindent
The two-step quantile regression estimator is computed using two different R \citep{R_software} packages. For the first step, the LQMM estimation method is implemented by
the \texttt{lqmm}() function from the package \texttt{lqmm}
\citep{JSSv057i13, geraci2014linear}. For the second step, we use  standard quantile
regression implemented by the function \texttt{rq}() from the
\texttt{quantreg} package \citep{quantreg_2020version}.
Bootstrap datasets are generated with
standard sampling functions.
An R function for the complete estimation
and inference process is available from the  corresponding author's
website.



\section{Simulations}\label{sec:simulation}

\subsection{Data generating model}


\noindent
We consider a data generating model inspired by the simulation designs in
\cite{koenker2004quantile} and \cite{geraci2014linear}. 
Specifically,
\begin{align}\label{eq:simmodel}
Y_{ij} = \beta_0 + \beta_1 x_{ij} + u_i + (1 + \gamma x_{ij})e_{ij}, \hspace{5mm} i=1,...,N, \hspace{2mm} j=1,...,n_i,
\end{align}
where $u_i\stackrel{iid}{\sim} N(0, \sigma_u^2)$,
$e_{ij}\stackrel{iid}{\sim} N(0, \sigma_e^2)$,  $x_{ij}$
are uniformly distributed on $(0,1)$ and $\gamma \geq 0$ is a
homoscedasticity-departure parameter. Notice that $1+\gamma x_{ij}$ is always
positive.  When $\gamma\neq 0$, the covariate has both a location
shift and a scale effect \citep{koenker2004quantile}. In the
homoscedastic case (i.e. $\gamma=0$), the correlation between
observations from the same cluster is
$\frac{\sigma_u^2}{\sigma_u^2 + \sigma_e^2}$. With a slight abuse of
notation, we refer to this ratio as the interclass correlation
coefficient (ICC) even when $\gamma>0$.

Model \eqref{eq:simmodel} implies the following quantile regression model
\begin{align}\label{eq:sim-cond-quantile}
Q_{Y_{ij}|x_{ij}, u_i} (\tau)= \beta_0^{\tau} + \beta_1^{\tau}x_{ij} +u_i,
\end{align}
where $\beta_0^{\tau} = \beta_0 + \sigma_e \Phi^{-1}(\tau)$ and
$\beta_1^{\tau} = \beta_1 + \gamma \sigma_e  \Phi^{-1}(\tau)$, with
$\Phi$ denoting the cumulative distribution function
for the $N(0,1)$ distribution.
In particular, the quantiles are of the same form as \eqref{eq:qregfull},
\textcolor{black}{with $X_{ij}=(1,x_{ij})$ and $Z_{ij}=1$,} 
and with $u_i^\tau$ not
depending on $\tau$. 
When $\gamma=0$ the slope parameter of the quantile is
constant across $\tau$, i.e., $\beta_1^{\tau}=\beta_{1}$, while the covariate effect differs between
quantile levels when $\gamma \neq 0$. Irrespective of the choice of $\gamma$, the regression parameter for the median,
$\beta_1^{0.5}$, does not depend on $\gamma$, since $\Phi^{-1}(0.5)=0.$    

  Notice that the data
generating model implies that the marginal-type quantile at level $\tau$ of $Y_{ij}$ given $x_{ij}$ (but not conditional on $u_i$)
is given by
\begin{align}\label{eq:marq-quantile}
 \beta_0 + \beta_1 x_{ij} + \Phi^{-1}(\tau)\sqrt{\sigma_u^2 + (1 + \gamma x_{ij})^2\sigma_e^2}.
\end{align}
In the heteroscedastic setting ($\gamma>0)$ this expression is not linear in $x_{ij}$, in contrast with
\eqref{eq:sim-cond-quantile}, and a linear approximation has
parameters that are different from $\beta_0^\tau$ and
$\beta_1^\tau$. This shows that a marginal estimation approach 
aims at
different parameters compared to those in \eqref{eq:sim-cond-quantile}.
\color{black}

We are going to compare our proposed estimators to the marginal
estimator and the other estimation methods discussed in Section~\ref{sec:estimation}.
To implement the approaches we use the function \texttt{rq}() of the
\texttt{quantreg} package \citep{quantreg_2020version} to perform standard quantile
regression and the \texttt{lqmm}() function of the
package \texttt{lqmm} \citep{JSSv057i13} to perform LQMM. More
specifically, we use Gauss-Hermite quadrature (option \texttt{lqmmType=''normal''}
in \texttt{lqmm}) with 15
quadrature points (\texttt{nK=15}) and derivative-free optimisation
(\texttt{lqmmMethod=''df''}).
Quantile regression with $\ell_1$ and 
$\ell_2$ penalization and cross validation for selection of the
penalty parameter is implemented in 
the function \texttt{cv.hqreg}() of the
\texttt{hqreg} package \citep{hqregRdoc}. We use five-fold cross validation.
Finally, we use $B=100$ bootstrap replications for bias-adjustment, where applicable.

\subsection{Comparison of estimation methods}\label{sec:compare}


\subsubsection*{Overall comparison for a benchmark scenario}
\noindent
In the model \eqref{eq:sim-cond-quantile}, we consider 
true (mean) parameters $\beta_0=\beta_1=1$, homoscedasticity departure
parameter $\gamma=0.4$, variances $\sigma_u^2=\sigma_e^2=1$, and thus
$\text{ICC}=0.5$. The main focus is on the quantile level $\tau =
0.1$ that is somewhat extreme; then true parameter values
amount to $\beta_0^\tau=-0.281$ and $\beta_1^\tau= 0.487$.
Define the ``benchmark scenario'' by the case with $N=500$ clusters of size
$n_i=6$ ($i=1,...,N$); we use this scenario to study the performance of the
estimators in the situation with $N \gg n_i$.  

\begin{figure}[htb!]
  \begin{center}
        \includegraphics[width=\linewidth]{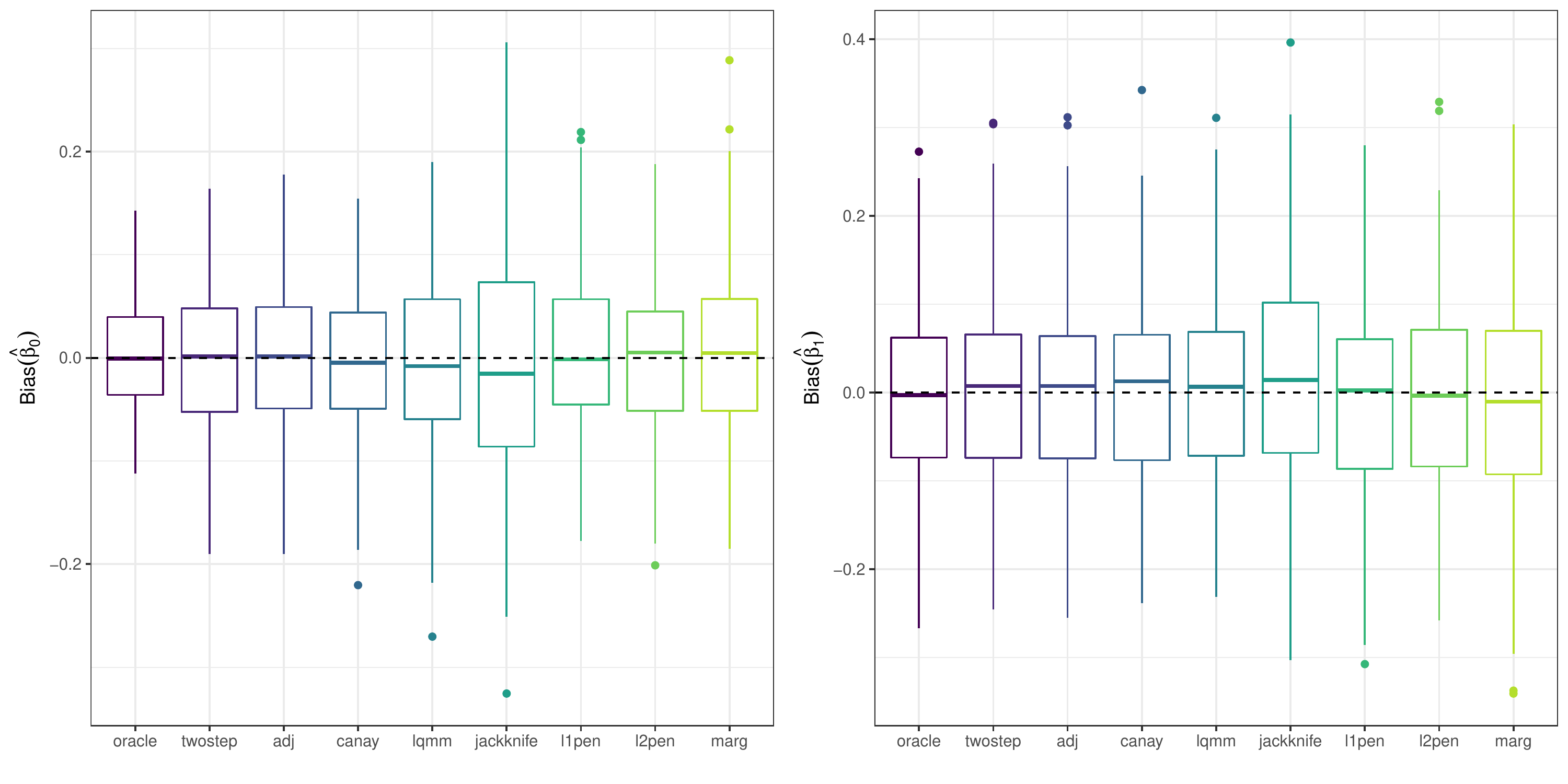}
    \includegraphics[width=\linewidth]{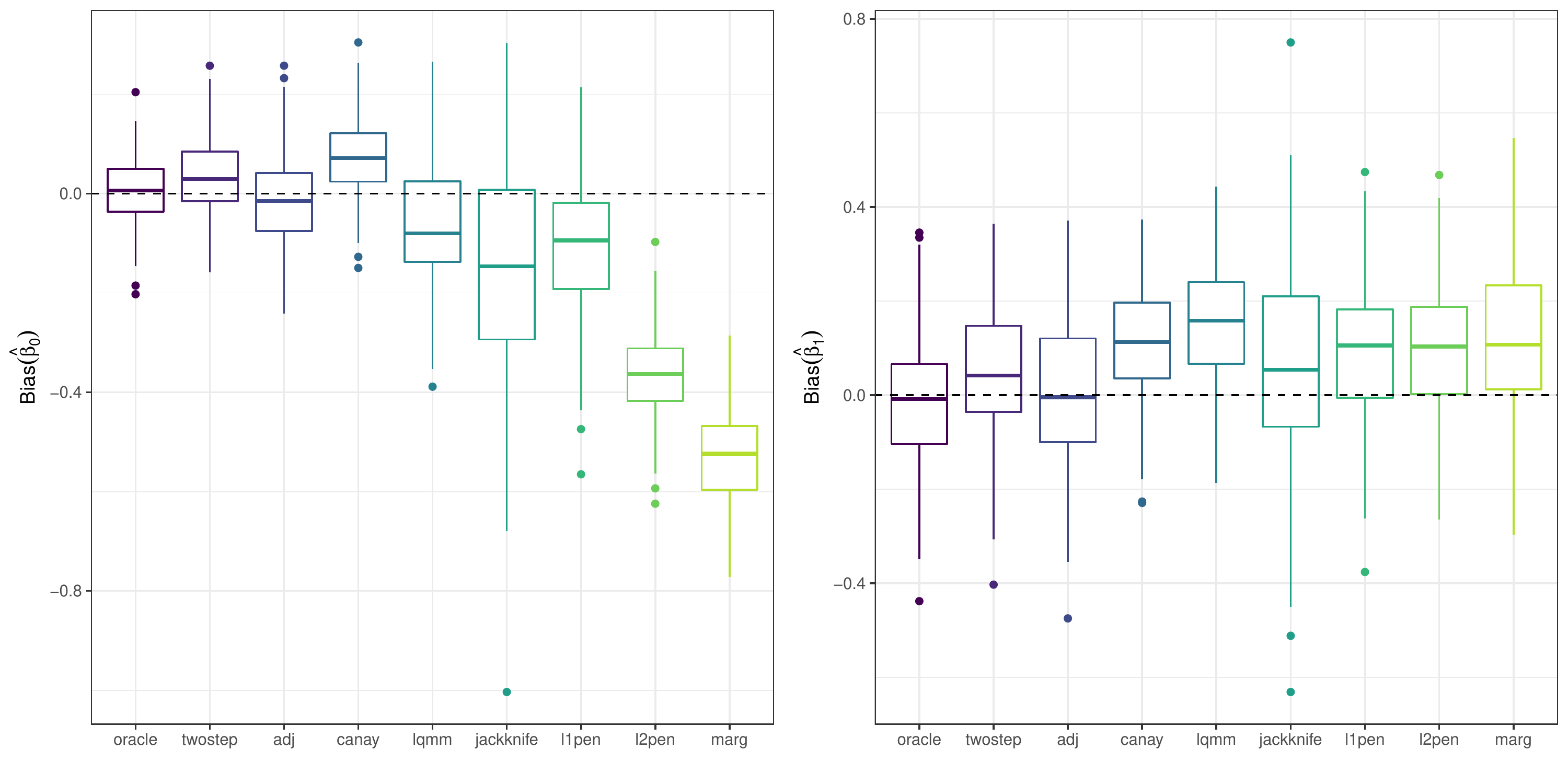}

		\caption{Bias for different estimators of $\beta_0^\tau$ (left) and $\beta_1^\tau$ (right) for 200
			datasets from the benchmark scenario. The
                        quantile level is 0.5 (top) and 0.1
                        (bottom). The true parameter values are 
                        $\beta_0^{0.5}=\beta_1^{0.5}=1$ 
                        and 
                      $\beta_0^{0.1}=-0.281$, $\beta_1^{0.1}=0.487$, respectively.}
                    \label{fig:plot1}
                    \end{center}
	\end{figure}

Figure
\ref{fig:plot1} shows the boxplots of the bias for $\beta_{0}^{\tau}$ (left) and
$\beta_{1}^{\tau}$ (right) corresponding to quantile levels $\tau=0.5$ (top) and $\tau=0.1$ (bottom), based on $200$ Monte Carlo simulations.
We compare the proposed two-step estimator and its adjusted version (\texttt{twostep} and
\texttt{adj}, respectively), the estimator from  \cite{doi:10.1111/j.1368-423X.2011.00349.x} (\texttt{canay}), the LQMM estimator (\texttt{lqmm}) and its jackknife-based adjustment (\texttt{jackknife}), the estimators arising from penalized quantile regression, both with $\ell_1$ and $\ell_2$ penalties (\texttt{l1pen} and \texttt{l2pen}, respectively), the marginal estimator arising from
standard quantile regression (\texttt{marg}), and the estimator from \eqref{eq:oracle}
where the actual random effects are used in the computations (\texttt{oracle}). The oracle
estimator is unfeasible in practice, but is used as a reference to study the effect of random effects being latent.

All nine estimators have similar distributions  for $\tau=0.5$, except
the jackknife-adjusted estimator, which has slightly larger variation for both
parameters.
The results are more interesting for $\tau=0.1$. 
Focusing first on the  methods developed in this paper, the unadjusted
two-step estimator has a smaller bias (component-wise) than the other estimators studied;
yet, there is still some bias left compared to the oracle estimator. 
The bias-adjusted estimator, on the other hand, has a very small
bias (for each component) and variance that is slightly larger than that of the
oracle estimator, but comparable to the other competitors.

The estimator proposed by \cite{doi:10.1111/j.1368-423X.2011.00349.x}
has a comparable bias to the other estimators when it comes to the
slope, but it shows positive (but small) bias for the
intercept. The variance is small for both components of the quantile regression parameters. 
Results for the LQMM estimators and the estimators from
\cite{koenker2004quantile} based on $\ell_1$ penalisation are similar
and show a small bias for both components. The estimator based on
$\ell_2$ penalisation has the same properties for the slope, but has a
larger bias for the intercept. The jackknife-based adjustment
of the LQMM estimator reduces the bias for the slope parameter, but not for the
intercept, and generally, it has large variation.

As expected, the standard quantile regression estimator, which completely ignores the cluster structure, leads to increased bias. 
The bias is particularly severe
for the intercept, whereas the bias for the slope is comparable to
that of Canay's estimator, the LQMM estimator, and the penalization-based
estimators. This is interesting, as it indicates that these latter 
estimators effectively estimate the slope coefficient in (a linearized
version of) a marginal quantile model rather than in the conditional quantile model.
%

Additional simulation results are included in the appendix;
Tables \ref{table:combi50_othermethods}--\ref{table:combi10} show results for settings where $(N, n_i)$ differ from the
benchmark scenario, and for quantile levels $\tau=0.1,0.5$. The
conclusions from Figure~\ref{fig:plot1} are confirmed; in particular an advantage of the proposed estimators
is observed for $\tau=0.1$ (Tables \ref{table:combi10_othermethods} and
\ref{table:combi10}).  In passing, we note that the $\ell_1$-penalized estimator is
preferable to the $\ell_2$-penalized estimator in all settings, and that
the jackknife estimator reduces bias for $\hat\beta^\tau_1$ but
increases bias for $\hat\beta^\tau_0$ and has larger variance. For those reasons we do not study the $\ell_2$-penalized and the
jackknife estimators any further. The remaining estimators are
discussed in more detail in the next section.

\color{black}

The average computing time per simulated dataset for the 
bootstrap-adjusted two-step estimator was 18.83 seconds. By comparison, the
computation time for the LQMM estimator was 0.15 seconds.  The
difference reflects the additional $B=100$ iterations involving LQMM estimation and the construction of the confidence intervals that are required by the proposed method. 
{The average
  computation time for Canay's estimator was 0.58 seconds.}
The average computation time for the $\ell_1$-penalized estimator was
72.29 seconds, partly due to the cross-validation step.  The
computation time for the $\ell_2$-penalized estimator was close to that
of the $\ell_1$-penalized, and computations for the jackknife
adjusted estimator took about three times longer than computations for
LQMM. {Computations were run on a commodity PC with 2.9 GHz Dual–Core Intel Core i5 processor 5287U.} 


\color{black}
\subsubsection*{Bias for LQMM, $\ell_1$-penalized, and Canay's estimator for extreme quantile levels}
\noindent
For quantile level 0.1, the bias of the LQMM, $\ell_1$-penalized, $\ell_2$-penalized and
Canay's estimators in the bottom of Figure~\ref{fig:plot1} is quite large.
This flaw is reported for Canay's estimator in a
  simulation study with $N$ much larger than $n_i$ and varying 
quantile levels \citep{doi:10.1111/j.1368-423X.2011.00349.x}; however, to
the best of our knowledge, the bias has not been documented thoroughly in
the literature for the other estimators.
The $\ell_1$-penalized estimation is carried out in \cite{koenker2004quantile} for 
a simulation model similar to ours, but only for the median ($\tau=0.5$)
where all estimators are unbiased. 
LQMM estimation is analyzed in \cite{geraci2014linear} in many
simulation scenarios with good overall performance,  but the
dependence on bias of sample size ($N$ and $n_i$) is not studied
in the presence of heteroscedasticity. 

\begin{figure}[hbt!]
  \begin{center}
		\includegraphics[width=\linewidth]{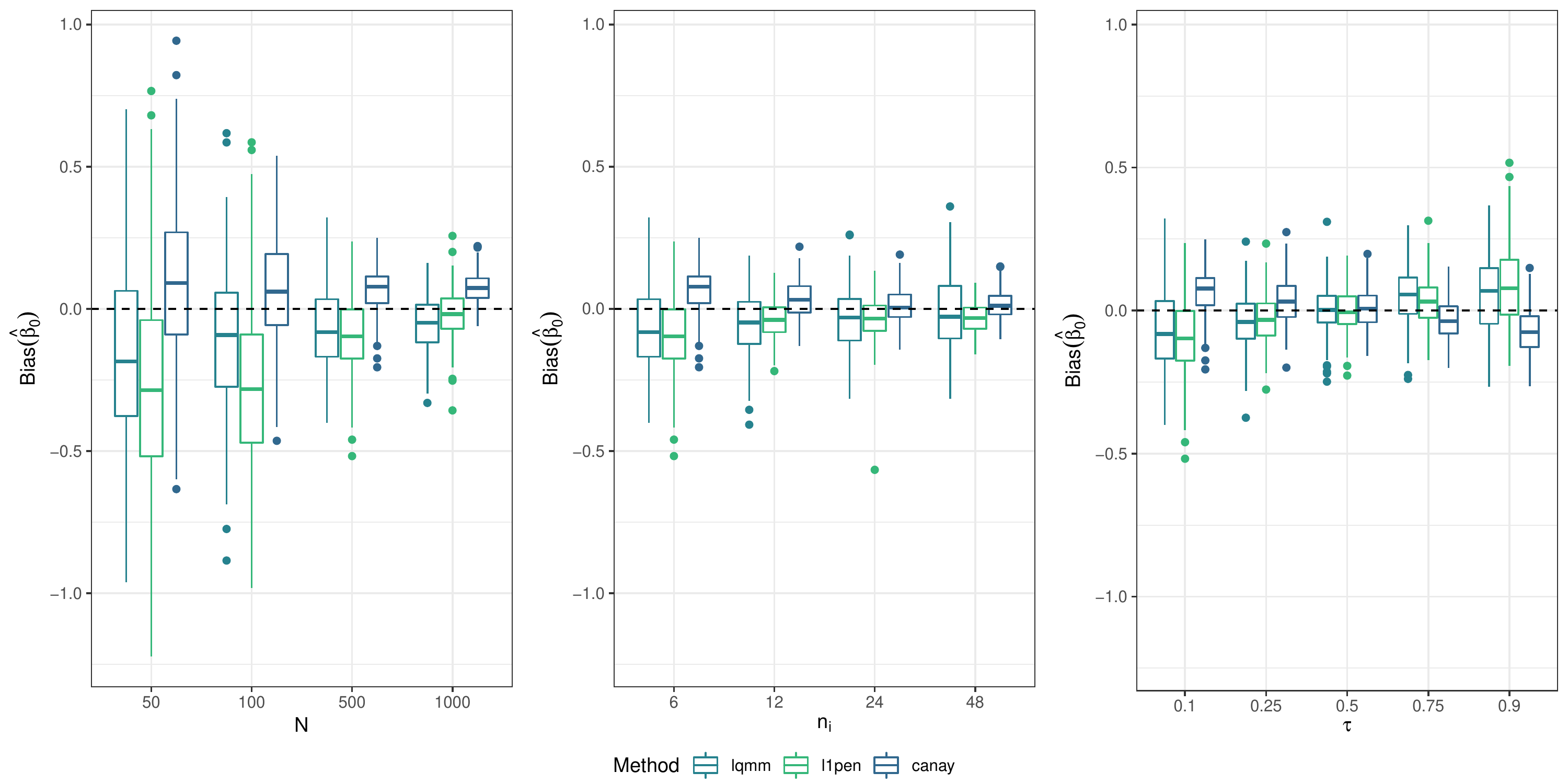}
		\includegraphics[width=\linewidth]{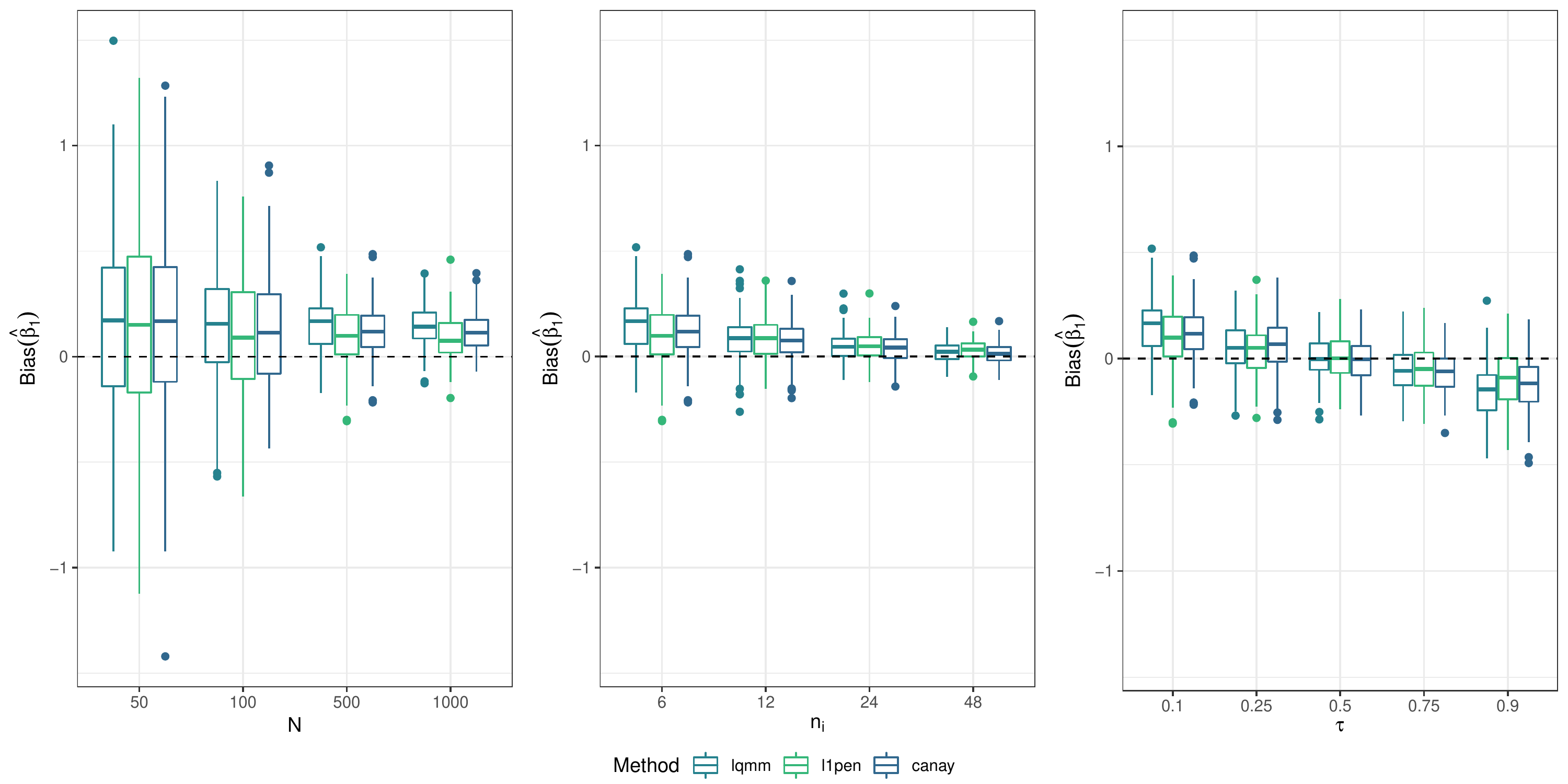}

		\caption{Boxplots for estimators of $\beta_0^\tau$ (top) and
			$\beta_1^\tau$ (bottom) for 200
			datasets from a selection of the traditional methods 
			with
			varying 
			$N$ (left panels), $n_i$ (middle panels) and $\tau$
			(right panels). The factors that do not vary are kept fixed at benchmark
			values: $N=500$, $n_i=6$, $\tau=0.1$.}
                      \label{fig:plot2}
\end{center}
	\end{figure}

Figure \ref{fig:plot2} shows boxplots of the bias for the LQMM, the $\ell_1$-penalized, and
Canay's estimator for various number of clusters, $N$, cluster sizes, $n_i$, and at different quantile levels, $\tau$; results are based on 200 replications.
We vary one factor at a time, while
keeping the others fixed at their benchmark values ($N=500$,
$n_i=6$, $\tau=0.1$). 
%
\color{black}
As a
consequence, the benchmark scenario appears in each
panel. The top plots show the results for the intercept, while the
bottom row shows results for the slope.

Generally, the magnitude of the bias decreases as the number of observations per cluster increases for fixed $N$ (central panels): this confirms the existing asymptotic results
\citep{koenker2004quantile, doi:10.1111/j.1368-423X.2011.00349.x}.
However, when the cluster size, $n_i$, is fixed (left most panels), there is non-negligible bias for these estimators,
as the sample size, $N$, increases. The results are valid for both parameter components, but in particular for the slope (bottom panel).
In other words, the
estimators are not consistent for $\beta_1^\tau$ in the asymptotic
scenario with a fixed (and small) number of repeated measurements and
increasing the number of clusters. The bias behavior is worse for quantile levels closer to the boundaries, $\tau=0.1$ or $\tau=  0.9$, than for levels closer to the median, $\tau=0.5$ (right panels).

The three methods are comparable for estimation of $\beta_1^{\tau}$
whereas there are subtle differences for $\beta_0^{\tau}$: LQMM and $\ell_1$-penalized estimators behave similarly, except for
small values of $N$; Canay's estimator has bias of opposite sign and
of smaller size as well as smaller variation compared to the two other
methods. 
%
Further simulation scenarios are presented in Tables \ref{table:combi50_othermethods} and
\ref{table:combi10_othermethods} in the appendix, showing similar results. 
\color{black}

\subsection{Performance of the proposed estimators}

\subsubsection*{Bias and variation}
\noindent
We now turn to a more detailed study of our proposed  
estimators. Figure \ref{fig:plot3} has the same structure as
Figure~\ref{fig:plot2}, but now includes the oracle estimator (as an
infeasible point of reference), the LQMM estimator (as a 
representative of the existing methods, cf.\ Figure~\ref{fig:plot2},
and as starting point of our two-step procedure), 
and the unadjusted and adjusted two-step estimators. 
{Results are based on 1000 replications. The benchmark
scenario ($N=500$, $n_i=6$, $\tau=0.1$) was also considered in
Figure~\ref{fig:plot1}, but notice that the results of Figure~\ref{fig:plot3} summarize performance in 1000 simulations, while only 200 simulations were considered in Figure~\ref{fig:plot1}, due to the increased computational burden required by some of the alternative methods.

\begin{figure}[h!]
  \begin{center}
		\includegraphics[width=\linewidth]{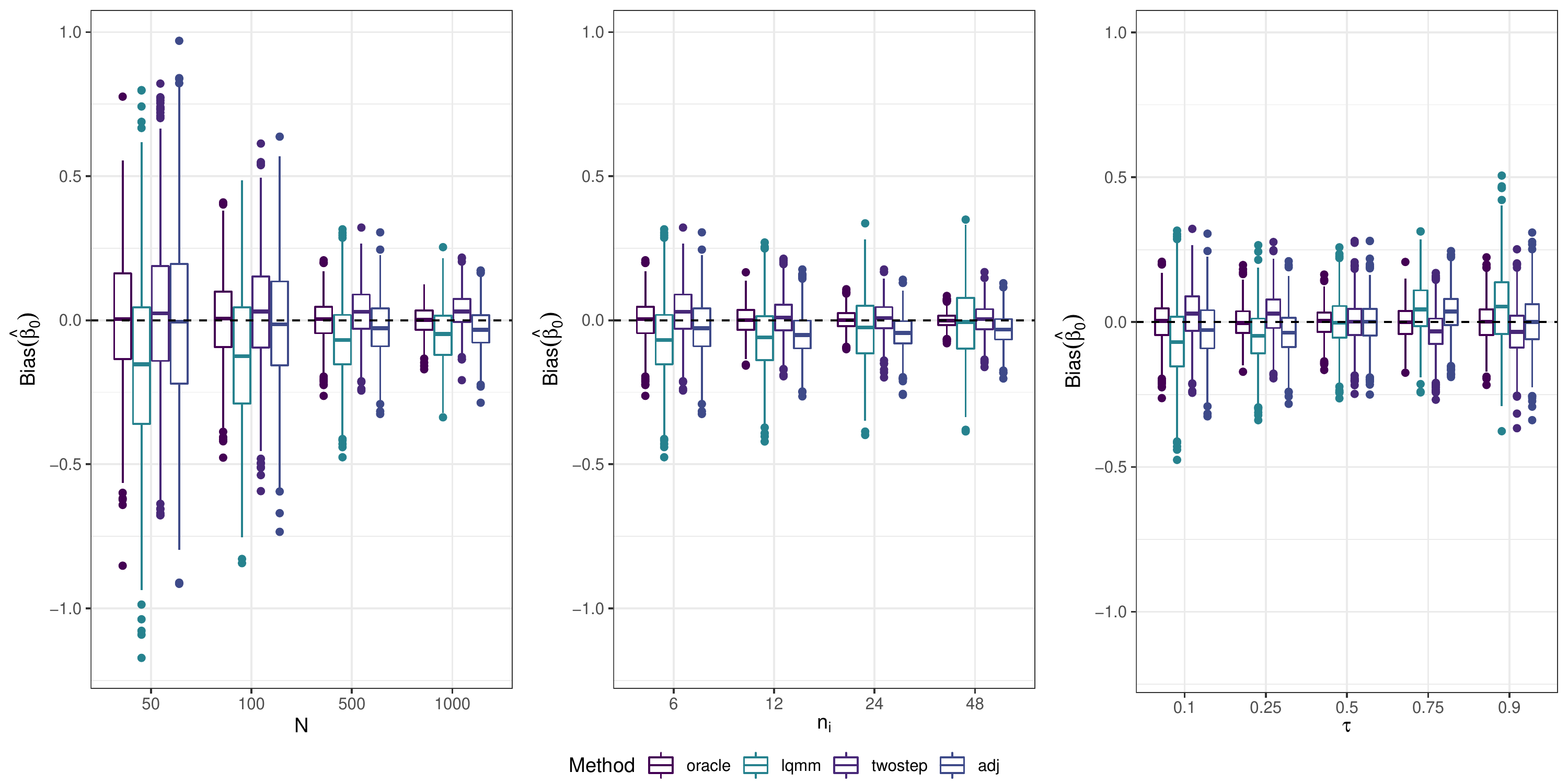}
	\includegraphics[width=\linewidth]{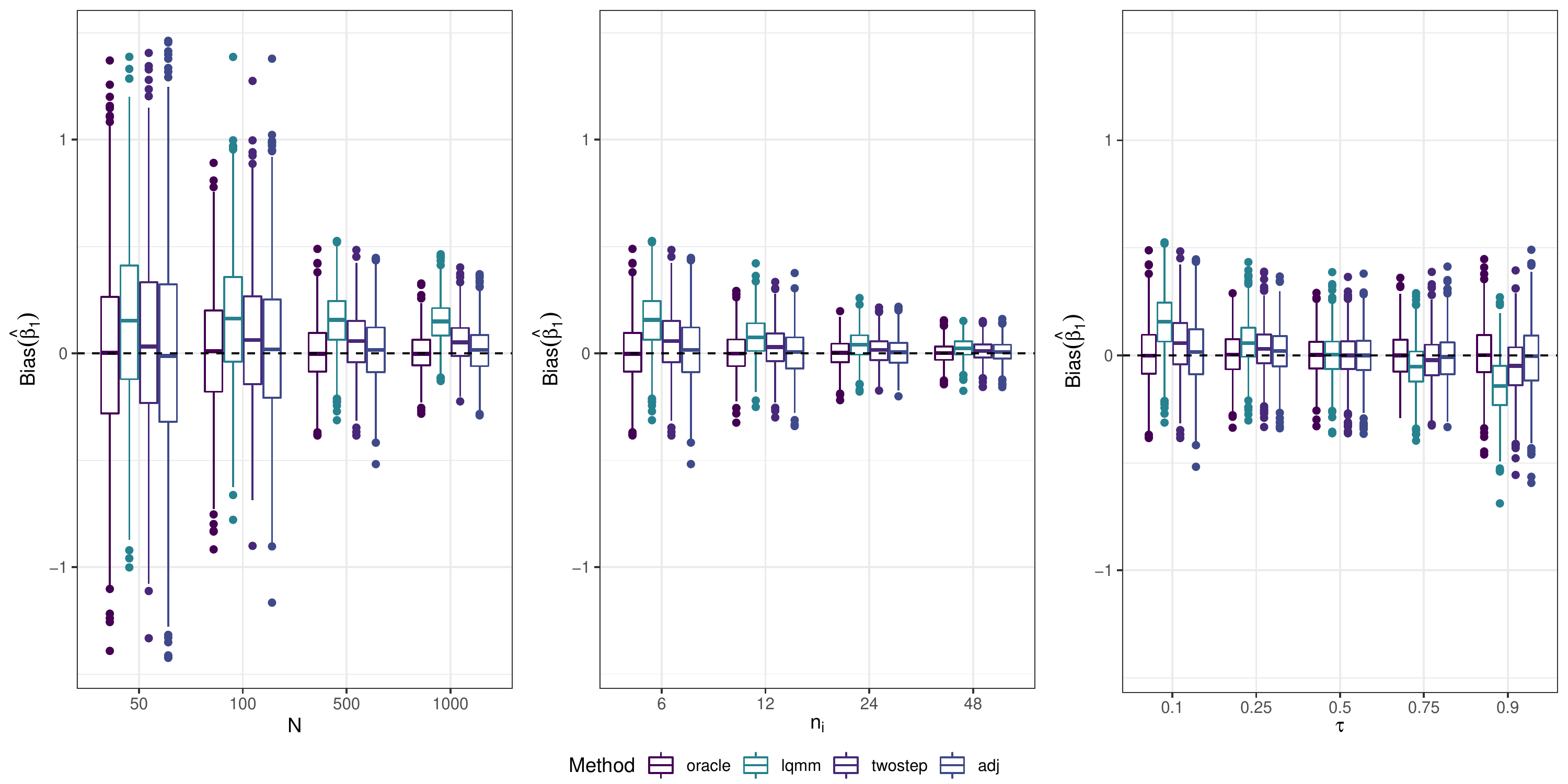}
		\caption{Boxplots for estimators of $\beta_0^\tau$ (top) and
			$\beta_1^\tau$ (bottom) for 1000
			datasets from the oracle estimator, the
			LQMM estimator and the adjusted two-step estimator with
			varying $N$ (left panels), $n_i$ (middle panels) and $\tau$
			(right panels). Parameters that do not vary are kept fixed at benchmark
			values: $N=500$, $n_i=6$, $\tau=0.1$.}
		\label{fig:plot3}
              \end{center}
              	\end{figure}

For the slope quantile regression parameter, $\beta^{\tau}_1$ (bottom panels), the bias is reduced for
the two-step estimator compared to the LQMM estimator and is
almost completely removed in all
scenarios for the bias-adjusted estimator. The variability is
  only slightly larger than the variability of the oracle estimator.
For the intercept quantile regression parameter, $\beta^{\tau}_0$, the bias is considerably reduced for the proposed two-step estimators compared to the LQMM estimator
{when the cluster size is small} (top left panel). For large clusters
the unadjusted two-step
estimator shows the best performance in terms of both bias and variance (top central panel).
\color{black}

\color{black}
Results for more combinations of $N$, $n_i$ and $\tau$ are reported in
the appendix. 
For the median, $\tau=0.5$ (Table \ref{table:combi50}), all three estimators are unbiased and show similar
variability.
For $\tau=0.1$ (Table
\ref{table:combi10}), the situation is more complex. Nonetheless, the
proposed two-step estimators (without adjustment) yields a smaller RMSE
than the LQMM counterpart. Consider the estimation of the slope
parameter $\beta^\tau_1$: all estimators seem to show similar
variability, however the two-step estimators indicate a considerably
improved bias behavior compared to the LQMM estimator.
The numerical
studies show that the cluster size has a larger impact on estimation
performance than the number of clusters; compare the RMSE when the
number of observations is kept fixed to say $3000$ composed by 1) $N=1000$ clusters of size $n_i=3$ and 2) $N=500$ clusters of size $n_i=6$.  
\color{black}


\color{black}

Figure~\ref{fig:plot4} compares the two-step estimators with the
oracle and LQMM for three extra scenarios that have 
larger heteroscedasticity ($\gamma=1$) or
larger within-cluster relative variance ($\sigma_u^2=1.5$,
$\sigma_e^2=0.5$ yielding ICC = 0.75), or larger total variation ($\sigma_u^2=\sigma_e^2=1.5$)
compared to the benchmark scenario. All other simulation parameters
are kept fixed to the values from the bechmark setting.
The changed parameter settings have larger impact on the distribution of the LQMM estimator than on the distribution
of the two-step estimators. In particular, 
the two-step
estimation results in improved bias performance compared to the LQMM
estimator, irrespective of the setting.

\begin{figure}[h!]
  \begin{center}
            		\includegraphics[width=\linewidth]{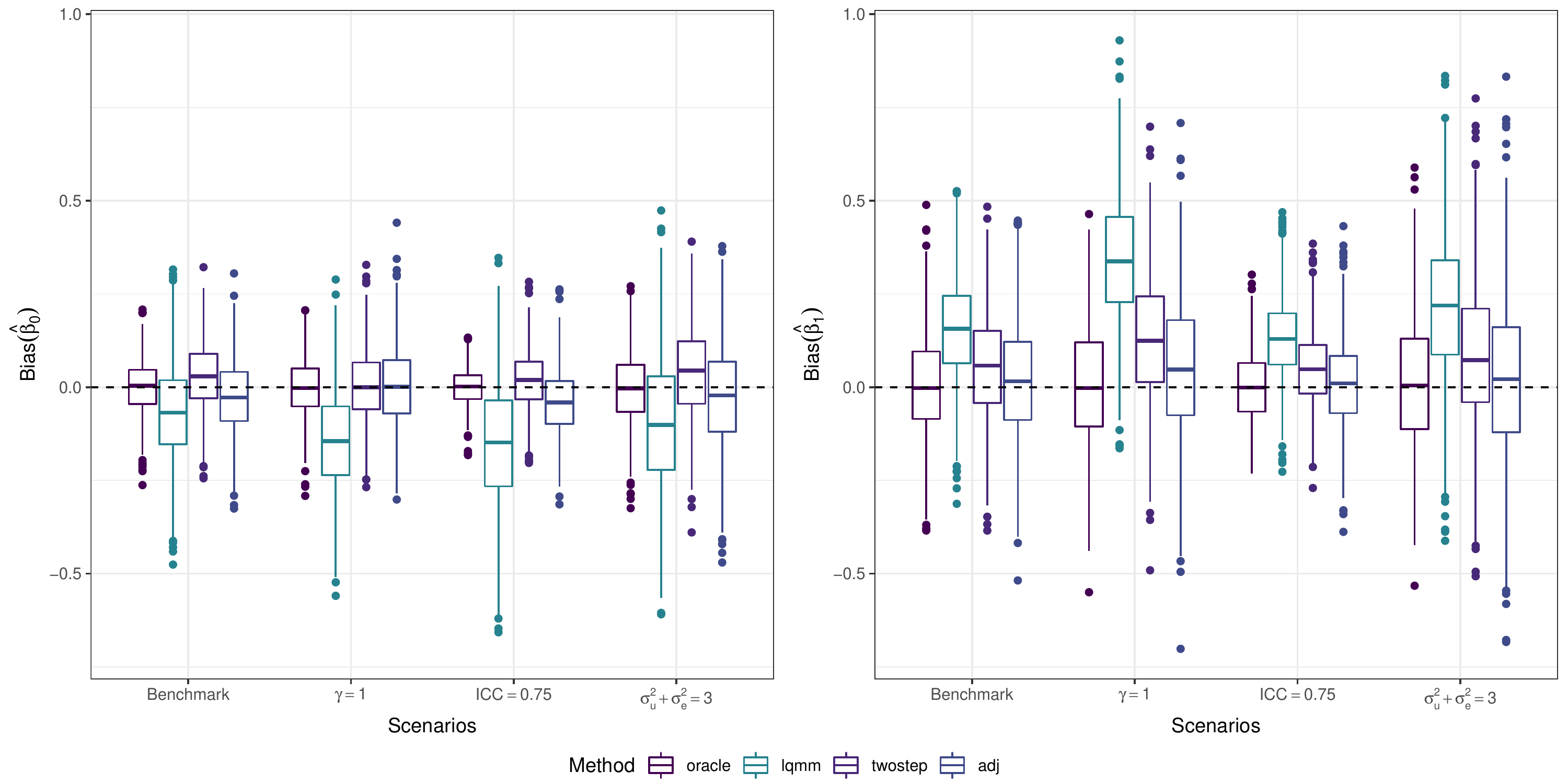}
		\caption{Boxplots of the estimates of $\beta_0^\tau$ (left)
                  and $\beta_1^\tau$ (right) obtained using oracle method, 
                  LQMM, and the two-step estimators with and without adjustment for 
                  two-step estimation for the benchmark scenario and
                  scenarios with larger homoscedasticity,
                  larger ICC, and larger variance. All the other simulation factors are kept constant to their values of the benchmark scenario.
                  Results are based on 200 simulations.}
		\label{fig:plot4}
                \end{center}
	\end{figure}

\subsubsection*{Confidence intervals and comparison of bootstrap strategies}

\noindent
%
Next, we turn to evaluating the proposed bootstrap scheme for decreasing the estimator's bias and construction of confidence intervals. 
We compare the proposed mixture of standard and wild resampling (denoted by RW) with other types of data resampling, with respect to bias-adjustment in estimating the parameters, as well as the actual coverage and average length of the confidence intervals.

\begin{description}

    \item[Resample random effects and residuals (RRR)]
      A bootstrap sample takes the form
      $\{(Y_{ij}^{*b}, X_{ij}, Z_{ij})_{j=1}^{n_i}, u_i^{\tau,*b}\}_{i=1}^N$ where
      $Y_{ij}^{*b} = {Z}_{ij}^T u_i^{\tau,*b} +
      {X}_{ij}^T\hat{\beta}^\tau_{\text{two-step}} +
      \varepsilon_{ij}^{*b}$, with $ \varepsilon_{ij}^{*b}$ obtained
      from a standard sampling with replacement procedure from the
      observed residuals, $\{ \varepsilon_{ij}
      \}_{i,j}$, and $u_i^{\tau,*b}$ is sampled from $\mathcal{U}$. In contrast to RW sampling, there is no coupling between
        covariates and residuals. 
\cite{Carpenter2003} has proposed the method for mean regression for multilevel
        data. 
       Notice that residuals could 
	also be sampled cluster-wise in order to maintain within-cluster
	dependence not accounted for by the random effect, but we do
        not consider this.

      \item[Resample clusters (RC)] The clusters are sampled with
      replacement in a completely non-parametric way. More specifically, $i_1^*, \ldots, i_N^*$ are sampled
	with replacement from $\{1,\ldots,N\}$, and a
	bootstrap dataset consists of
	$
	(Y^*_{ij}, X^*_{ij}, Z^*_{ij} ) =
        (Y_{i^*_i j}, X_{i^*_i j}, Z_{i^*_i j} )$, $i=1,\ldots,N, \ j=1,\ldots,n_i$. 
        Within-cluster dependence is maintained because complete
        clusters are sampled. 
	The method, also known in the literature as \emph{cross-sectional resampling}
	\citep{galvao2015bootstrap}, is used by
        \cite{doi:10.1111/j.1368-423X.2011.00349.x} and
        \cite{geraci2014linear} to construct confidence
        intervals. {\cite{Karlsson2009} uses RC in an attempt to correct
        for estimation bias in a nonlinear quantile regression for
        longitudinal data, using a marginal perspective, but experienced limited gain.}

      \item[Cluster-wise wild bootstrap (CW)] The idea is to use wild
      bootstrap for the sum of random effects and error
      terms.  Specifically, let   $r_{ij} = Y_{ij} - X^T_{ij}\hat \beta_{\text{two-step}}^\tau$ be the residuals corresponding to the two-step estimation, and let $w_i$s be a random  sample from \eqref{eq:w_distr}.
            The bootstrap sample is $\{(Y_{ij}^{*b}, X_{ij}, Z_{ij})_{j=1}^{n_i}\}_{i=1}^N$, where  $ Y_{ij}^{*b} = X^T_{ij}\hat
        \beta_{\text{two-step}}^{\tau} + w_i |r_{ij}|$.
        In contrast to the residuals $\varepsilon_{ij}$ used
        for RW, $r_{ij}$ are defined without subtraction of predicted random
        effects (often referred to as ``level zero residuals''). 
        Also, same weight $w_i$ is used for all the observations within cluster $i$ in
        order to preserve dependence within clusters. 
	This resampling scheme is used by \cite{ModugnoGian2015} in
        the context of multilevel models for mean regression, but does not
        appear to have been used for quantile regression.
	
      \end{description}

The RW and RRR sampling schemes use bootstrap to approximate the joint
distribution of $(u_i^\tau, Y_{ij})$, whereas the other two bootstrap
methods approximate the distribution of $Y_{ij}$ only.
As RC- and CW-based approaches do not involve generation of random effects,
SE-adjustment confidence intervals are only applicable for RW and
RRR. {The bias-adjusted estimator and basic confidence intervals, on
the other hand, can be computed for any of the four bootstrap schemes.}

Table~\ref{table:coverage} shows bias and actual coverage rates for
confidence intervals with an intended level of 95\%.  We employ the benchmark scenario, except for a varying number of
clusters (same simulated data as in the left part of Figure~\ref{fig:plot3}).
Results are based on 1000 simulated datasets. 
SE-adjustment confidence intervals generated with the RW bootstrap
method give the best coverage rates, close to the nominal 95\% in all
scenarios. Basic confidence intervals with RW bootstrap are also good for 
$\beta^\tau_0$ when $N$ is large, whereas coverage rates are below 0.90 for $\beta^\tau_1$.
RRR and RW produce similar coverage rates for $\beta^\tau_1$, but
no bias-adjustment is obtained with RRR (bias is equivalent to bias for the unadjusted
two-step estimator, not reported).  For $\beta^\tau_0$ the
coverage rates are slightly smaller for RRR compared to RW. 
{As expected, bootstrap method RC gives no bias reduction, neither for
$\beta^\tau_0$ nor $\beta^\tau_1$}, and coverage rates are
consequently never above 0.90.
The CW bootstrap method has poor coverage rates. For
$\beta^\tau_1$ the main reason is that the adjusted estimator has
large variability which is not properly taken into account, whereas
the explanation for $\beta^\tau_0$ is that CW introduces a large
bias such {that the confidence interval is located far from the true value.}

\begin{table}
	\centering
	\begin{tabular}{clcccccccc}
		& & \multicolumn{4}{c}{$\beta_0^\tau$} & \multicolumn{4}{c}{$\beta_1^\tau$} \\
		$N$ & & RW & RRR &RC & CW &RW & RRR&RC&CW\\ \hline
		& Bias & -0.01 & -0.03 & 0.03 &  -0.94&0.01 & 0.05& 0.05 &-0.03 \\
		$50$& Coverage, basic &0.87& 0.90& 0.88& 0.10&0.86& 0.90& 0.86& 0.42 \\
		& Coverage, SE-adj. &0.95&0.93&---&---& 0.95 &0.96&---& ---\\
		\hline
		
		& Bias & -0.02 &-0.03&  0.03& -0.9&0.03 &0.07 &  0.07& $<0.01$ \\
		$100$& Coverage, basic&0.89& 0.90& 0.90& 0.02&0.88& 0.90& 0.89& 0.39\\
		& Coverage, SE-adj.&0.95&0.92&---&---&0.94&0.95&---&---\\
		\hline

          	& Bias & -0.02 &-0.03 &  0.03 &-0.92&0.02 & 0.06 &  0.06 &-0.03\\
		$500$& Coverage, basic &0.94& 0.90& 0.90& $<0.01$&0.89& 0.89& 0.88& 0.36\\
		& Coverage, SE-adj.& 0.96&0.91&---&---&0.93&0.92&---&---\\
		\hline

          	& Bias & -0.03&-0.02 &  0.03&-0.91&0.02&0.05 &  0.05& -0.04 \\
		$1000$& Coverage, basic&0.94& 0.88& 0.86& $<0.01$&0.89& 0.88& 0.89& 0.31\\
		& Coverage, SE-adj.&0.95&0.89&---&---&0.93&0.90&---&---\\
		\hline
	\end{tabular}
	\caption{Bias and coverage rates of 95\% confidence intervals for the adjusted two-step method for different bootstrap
          schemes (RW, CW, RC, RRR) for 1000
			datasets. Basic confidence intervals are
          used for all bootstrap schemes, whereas SE-adjusted
          confidence intervals are only defined for RW and
          RRR. Cluster size is fixed at $n_i=6$ and the quantile level
          is $\tau=0.1$.}
          	\label{table:coverage}
      \end{table}

In summary, the semi-parametric bootstrap sampling methods using the
additive model structure for the quantiles (RW and RRR) with SE-adjusted confidence intervals show the best coverage 
properties. Nonetheless, the proposed two-step with RW-based adjustment results in the greatest bias reduction. 

\cite{geraci2014linear} and  \cite{doi:10.1111/j.1368-423X.2011.00349.x} use RC bootstrap for
construction of confidence intervals (Canay also uses asymptotic results),
and Table~\ref{table:coverage_compare} compares coverage rates and average lengths for their
confidence intervals and our SE-adjusted confidence intervals
based on RW sampling. The simulated data are the same as those used
for Table~\ref{table:coverage}. \cite{geraci2014linear} and
\cite{doi:10.1111/j.1368-423X.2011.00349.x} present estimation and
inference results regarding different settings than the ones
considered here, but our results are well in line with theirs. The LQMM and
Canay confidence intervals loose coverage for large $N$ because the
estimators are biased. For small $N$ the coverage is close to the
nominal level (bias plays a minor role because variation is large),
and the confidence intervals are shorter than those based on 
SE-adjustment, most likely because extra variability is introduced with the bias
adjustment.

\begin{table}
	\centering
	\begin{tabular}{clcccccc}
			& & \multicolumn{3}{c}{$\beta_0^\tau$} & \multicolumn{3}{c}{$\beta_1^\tau$} \\
		$N$ & & adj (RW) & lqmm & Canay & adj (RW)& lqmm & Canay \\ \hline
		& Bias & -0.01 & -0.16 & 0.07 & 0.01& 0.15& 0.13 \\
		$50$& Coverage &0.95 & 0.93& 0.93&0.95 &0.94& 0.93\\
		& Av. Length & 1.29 & 1.21 &  0.98 & 2.11 &1.67 & 1.55 \\
		\hline
		
		& Bias & -0.02 &-0.13&  0.07& 0.03 &0.17&0.13 \\
		$100$& Coverage& 0.95  & 0.90 & 0.93 &0.94 &0.93& 0.91\\
		& Av. Lengtvh &  0.88 & 0.93 & 0.70 & 1.38 & 1.22& 1.09 \\
		\hline

          	& Bias & -0.02 & -0.07&  0.07 & 0.02 & 0.15 & 0.13 \\
		$500$& Coverage &  0.96 & 0.90 & 0.84& 0.93 & 0.83& 0.81\\
		& Av. Length & 0.42 & 0.52 & 0.31 & 0.56 &0.57 &  0.48\\
		\hline

          	& Bias & -0.03 & -0.05 & 0.07 & 0.02 &0.15&  0.13 \\
		$1000$& Coverage &0.95 & 0.90 & 0.73 & 0.93&0.70& 0.69\\
		& Av. Length & 0.30 & 0.40 & 0.22 & 0.38 &0.41 & 0.34 \\
		\hline
	\end{tabular}
	\caption{Bias, coverage rates of 95\% confidence intervals and average length of confidence intervals 
          for our adjusted two-step method as well as LQMM and Canay's methods for 1000
			datasets. Cluster size is fixed at $n_i=6$ and the quantile level
          is $\tau=0.1$.
        }
          	\label{table:coverage_compare}
      \end{table}

\color{black}

\subsection{Additional simulation studies}
\noindent
At the suggestion of an anonymous reviewer, we further investigate the proposed method when the errors $e_{ij}$ are generated from a non-Gaussian distributions. Specifically, we use a scaled $t_3$-distribution and an
$ALD(0,\sigma_0,\tau_0)$ with $\tau_0=0.1$ and
$\sigma_0=
\frac{(1-\tau_0)\tau_0}{\sqrt{1-2\tau_0+2\tau_0^2}}=0.09939$.
Both distributions are scaled to have unit variance in order to make fair the
comparison with the standard normal errors scenarios considered previously. 
When
sampling from the ALD distribution, we consider both the
benchmark scenario and a departure from it, corresponding to
$\gamma=0$. Notice that 
the true values of $\beta_0^\tau$
and $\beta_1^\tau$ change compared to the standard normal case.
The results are shown in Table \ref{table:t3+ald} in the appendix
and should be compared to the relevant scenarios in
Table~\ref{table:combi10}. 

In the case of scaled $t$-distributed errors, the bias is reduced for the two-step
estimator, compared to the LQMM estimator, but it is not completely removed. 
The RW bootstrap correction reduces the bias even further
for $\beta_1^\tau$, but surprisingly it increases the bias for
$\beta_0^\tau$. This may be due to the inflated residuals that are
obtained with the wild bootstrap scheme, as they can
  be large in the situation of heavy-tailed errors, and
therefore have large impact on the estimation of bias for the
intercept.

In the case of heteroscedastic ALD errors ($\gamma > 0$), the bias of the LQMM
estimator for $\beta_1^\tau$ is reduced considerably compared to
the Gaussian case (Table~\ref{table:combi10}). The estimators' variability is also reduced in this setting, in spite of the error variance remaining fixed, because quantiles are generally estimated with
higher precision when the model is ALD than when it is Gaussian.  The two-step
estimator and the adjusted two-step estimator have almost the same
distributions as the LQMM estimator. For estimating the intercept, the performance of the proposed estimators
is superior to that of the LQMM, in terms of reduced bias and variability. 

When the errors come from a homoscedastic ALD ($\gamma=0$), the working distribution
for the LQMM estimation approach coincides with the data generating
mechanism. As expected, the LQMM
estimator of $\beta_1^\tau$ has a very good performance: no bias and small variance. The two-step estimators are also unbiased, but have slightly larger variance.
For estimating the intercept parameter, surprisingly, the LQMM
estimator shows a behavior comparable to the
heteroscedastic ALD case; in contrast the two-step
estimators have a much smaller bias and variance.

%

Finally, we also consider a quantile regression model involving both a random intercept and a random
slope. To be specific, the data are generated from the model $Y_{ij} = \beta_0 +u_i +  (\beta_1+v_i)x_{ij}   + (1+\gamma x_{ij}) e_{ij}$, where $u_i$ is generated as described in \eqref{eq:simmodel} and $v_i \stackrel{iid}{\sim}N(0, \sigma_v^2)$.
Out of the existing methods, only LQMM allows to incorporate random
slopes in the quantile regression; thus we compare the results of the
two-step estimation with LQMM solely.  Table \ref{table:rsl} shows the
results. \textcolor{black}{We see that irrespective of the
  sample size or cluster size, the two-step estimation without
  adjustment improves or maintains the RMSE compared to LQMM
  estimation. The adjusted
  two-step estimator generally shows the smallest bias, but at the expense of
  increased variability; for the estimation of the intercept parameter
  in the case of $n_i=12$ the unadjusted two-step estimator has the
  smallest bias and variance.}


%

\color{black}

\begin{table}[h]
	\centering
	\begin{tabular}{cclcccccc}
		&& & \multicolumn{3}{c}{$\beta_0^\tau$} & \multicolumn{3}{c}{$\beta_1^\tau$} \\
		$N$&$n_i$ & & lqmm & two-step & adj (RW) & lqmm & two-step & adj (RW)\\ \hline
		&& Bias & -0.03&-0.01 & 0.00 &0.15  &0.14  &0.04 \\
		$500$& 6& SD &0.13 & 0.09&0.12 &0.18 &0.17 &0.24  \\
		&& RMSE & 0.14 & 0.09 &0.12 &0.24 &0.22 &0.24 \\
		\hline
		
		&& Bias & -0.03 &-0.02 &-0.02 & 0.18 &0.17 &0.07 \\
		$1000$&6& SD & 0.08 &0.08 & 0.11 & 0.15&0.15 &0.21 \\
		&& RMSE &0.09  &0.08 & 0.11 & 0.23& 0.22&0.22 \\
          \hline

          		&& Bias & -0.07 & -0.02&-0.05 &0.06 &0.10 &0.04  \\
		$500$&12& SD& 0.12&0.07 &0.08&  0.14&0.11 &0.13 \\
		&& RMSE & 0.14&0.07 &0.10 & 0.15 &0.15 &0.14 \\
		\hline

	\end{tabular}
	\caption{Bias, standard deviation, and RMSE for the LQMM
		estimator (lqmm), the two-step estimator
		(two-step), and bootstrap-adjusted two-step estimator (adj)
		where bootstrap samples are generated with the RW method,
		and we consider \textcolor{black}{the model with random intercept as
                well as random slope.} 
		The quantile level is $\tau=0.1$, and results are from 200 replications.}
	\label{table:rsl}
\end{table}


\section{Data application} 
\label{sec:application}
\noindent

\noindent
AIDS Clinical Trial Group (ACTG) Study 193A \citep{PMID:9833742} is a randomized and double-blinded study of
patients affected by AIDS at severe immune suppression stage, with CD4 counts of less than 50 cells/mm$^3$. 
There are 1309 patients, who were assigned to one of four treatments, namely: 600 mg of zidovudine daily alternating monthly with 400 mg of didanosine (double treatment 1); 600 mg of zidovudine as well as 2.25 mg of zalcitabine,   both daily (double treatment 2); 600 mg of zidovudine as well as 400 mg of didanosine, both daily (double treatment 3); the combination of 600 mg of zidovudine, 400 mg of didanosine and 400 mg of nevirapine, all of them daily (triple treatment). The CD4 counts were recorded at a baseline visit and at the follow-up visits during the subsequent 40 weeks. The measurements were intended to be taken every eight weeks, but occasionally there were dropouts or skipped medical appointments; see Figure \ref{fig:data_visualisation}.
After excluding the subjects with a single measurement (baseline), there are $N=1187$ subjects remaining in the study; their number of repeated measurements, $n_i$, varies between two and nine with a median of four.
The data has been previously used as an illustrative application for mean regression frameworks in \cite{fitzmaurice2012applied} and it is available at the associated webpage (\href{https://content.sph.harvard.edu/fitzmaur/ala2e/}{https://content.sph.harvard.edu/fitzmaur/ala2e/}).

\begin{figure}[h!]
  \begin{center}
           \includegraphics[width=\linewidth]{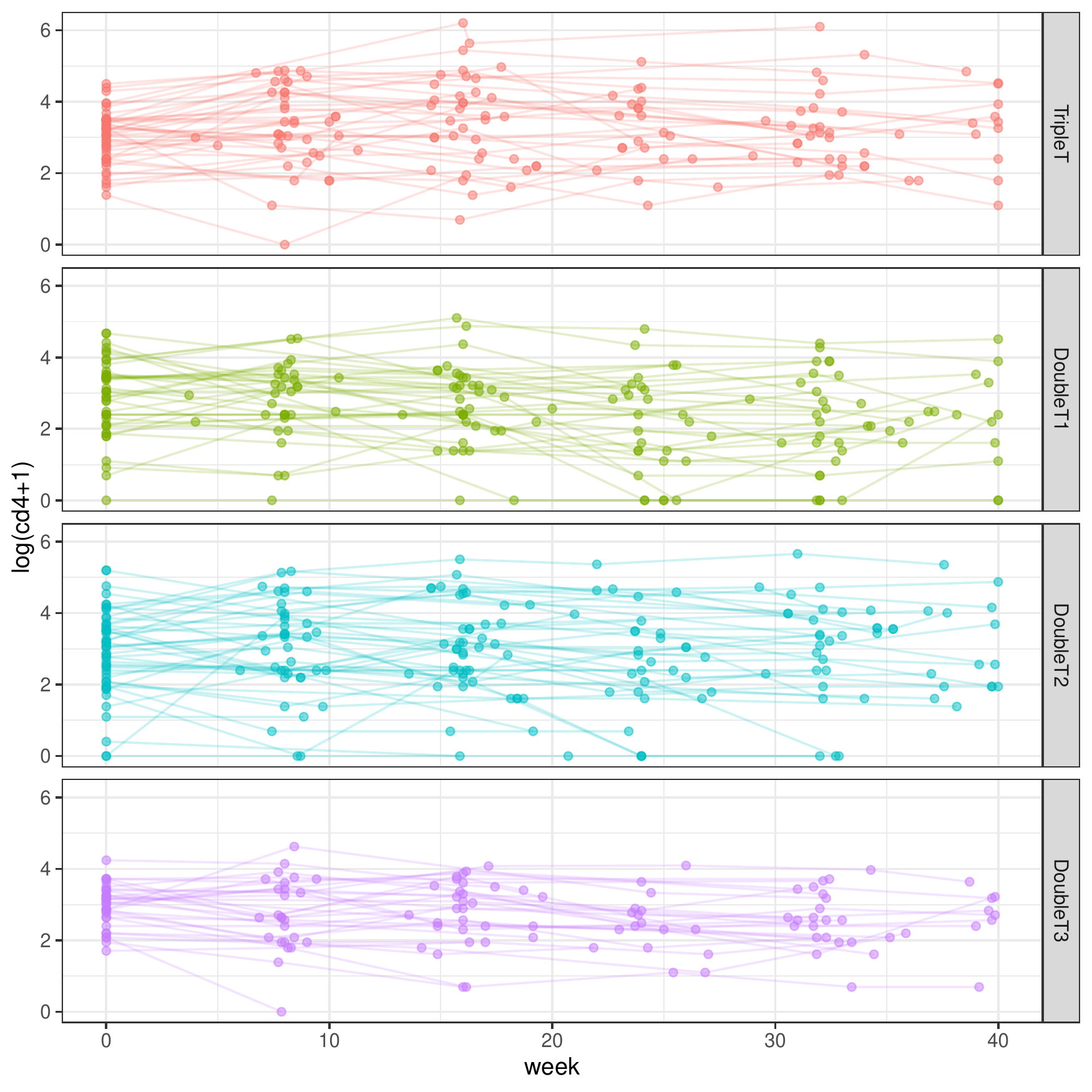}
		\caption{\textcolor{black}{Transformed CD4 counts for 200 patients, showing the records of 50 random subjects from each of the four treatment groups.} Observations from the same patients are connected with lines.}
                    \label{fig:data_visualisation}
                    \end{center}
	\end{figure}

        Our aim is to study the progression of the infection under the four treatment regimes for patients at different stages of immune suppression. Since CD4 counts are proxies for the stage of suppression---with lower CD4 counts corresponding to later stages---this can be obtained by studying the time trend for each treatment at different quantile levels. More specifically, an effective treatment reduces the decrease in CD4 counts, yielding a time trend closer to zero than a less effective treatment, and the effect may be different for early-stage patients (corresponding to high quantile  levels) than late-state patients (corresponding to low quantile levels). 
Figure~\ref{fig:data_visualisation} shows that subjects tend to have low or high CD4 counts
throughout, suggesting incorporation of subject-specific intercepts in
the model.

        As it is common in the literature, we log-transform the observed values and denote by $Y_{ij}$ the $\log(\text{CD4 count}+1)$ for patient $i$ at the $j$th hospital visit and by $t_{ij}$ the time of the $j$th visit, which is recorded by the number of weeks since the patient's baseline visit. We use dummy variables $\text{Treat}_{h}$ ($h=1,\ldots,4$) to indicate the assigned treatment, where $\text{Treat}_{1}$ corresponds to the triple therapy,  and $\text{Treat}_{2}$, $\text{Treat}_{3}$ and $\text{Treat}_{4}$ correspond to the three double treatments.
%
We account for age at baseline (variable $\text{Age}$) and sex (variable $\text{Sex}$, zero for females and one for males) as well. For simplicity of notation, we collect covariates relative to the $i$th patient at the $j$th follow-up visits into $X_{ij}$ such that $X^T_{ij} = (\text{Treat}_{1,i}, \text{Treat}_{2,i}, \text{Treat}_{3,i}, \text{Treat}_{4,i}, \text{Age}_i, \text{Sex}_i, t_{ij})$. To study the time-varying effect of treatment at quantile level $\tau$ of the response, let  $u_i^\tau$  be a subject-specific random effect associated with the quantile level $\tau$ and posit the following linear quantile regression model: 
\begin{equation}
\label{eq:model_separate_treats}
Q_{Y_{ij}|X_{ij}, u^\tau_i}(\tau) =  \sum_{h=1}^4 \beta_{0,h}^\tau \cdot \text{Treat}_{h,i} + \sum_{h=1}^4 \beta_{1,h}^\tau \cdot \text{Treat}_{h,i}\cdot t_{ij} + \beta_2^\tau  \cdot \text{Age}_i + \beta_3^\tau  \cdot \text{Sex}_i  + u_i^\tau.
\end{equation}

The slope parameters $\beta^\tau_{1,1},\ldots,\beta^\tau_{1,4}$ describe the behavior of CD4 counts over time, conditional on subject, and represent the main object of interest. As our interest is in the time varying effect of each treatment we are using the so-called ``explicit parameterization''; as a result, the model specification does not require a common intercept parameter. Estimation and inference are carried out using the proposed two-step estimation with adjustment; the results are compared with LQMM.

The estimated slope parameters for each treatment in part are plotted in Figure~\ref{fig:estimated_treatments_effects} for varying quantile levels. 
The left panels show the two-step estimates with adjustment and the corresponding 95\% confidence intervals for quantile levels $\tau\in \{0.1,0.15,\ldots,0.9\}$ (separate analyses). We used 100 RW bootstrap samples for the computations. 
The top panels concern the triple treatment: since the confidence band, corresponding to the two-step estimator, includes zero at all the quantile levels, it indicates that this therapy maintains an almost constant CD4 count during the study for subjects at any stage of their condition. For the other three treatments the situation is different. As depicted in the remaining panels, the two-step estimated coefficients $\hat\beta_{1,2}^\tau$, $\hat\beta_{1,3}^\tau$ and $\hat\beta_{1,4}^\tau$ are negative and significant at all the quantile levels, indicating that patients treated with either one of the double therapies must expect to see their CD4 count decrease over time.
Notice that there is a slight increase in the estimated $\hat\beta_{1,2}^\tau$ over quantile levels, which indicates that double treatment 1 makes the CD4 counts decrease {faster} for patients in the most severe conditions (lower quantile levels), whereas double treatments 2 and 3 appear to have more homogenous effects across patient groups.

In order to compare the treatments more directly we consider contrasts of the form $\hat\beta_{1,h}^\tau-\hat\beta_{1,1}^\tau$, which describe the difference in the effects between each double treatment and the triple treatment at quantile level $\tau$. 
The middle panels in Figure~\ref{fig:estimated_treatments_effects} show the estimated contrasts and the corresponding 95\% confidence intervals. Except for a single quantile level for double treatment 3, confidence intervals exclude zero, showing that the triple therapy is the most efficient treatment for patients in all infection stages. \cite{fitzmaurice2012applied} reported similar results for the mean. 

For comparison, the LQMM estimates and confidence intervals for the contrasts are shown in the right panels of Figure~\ref{fig:estimated_treatments_effects}.
Confidence intervals are based on 100 RC bootstrap samples. LQMM estimates are in the same range as the adjusted two-step estimates, albeit in general closer to zero. Moreover, the confidence bands are much wider, implying that the LQMM method does not find evidence for significant treatment differences for double treatments 2 and 3. 
This should not be surprising, since our numerical investigation showed that LQMM confidence intervals are wider (and coverage lower) than those corresponding to the adjusted two-step estimator, when the number of subjects is much larger than the number of repeated measurements; recall Table \ref{table:coverage}.

\begin{figure}[h!]
  \begin{center}
           \includegraphics[width=\linewidth]{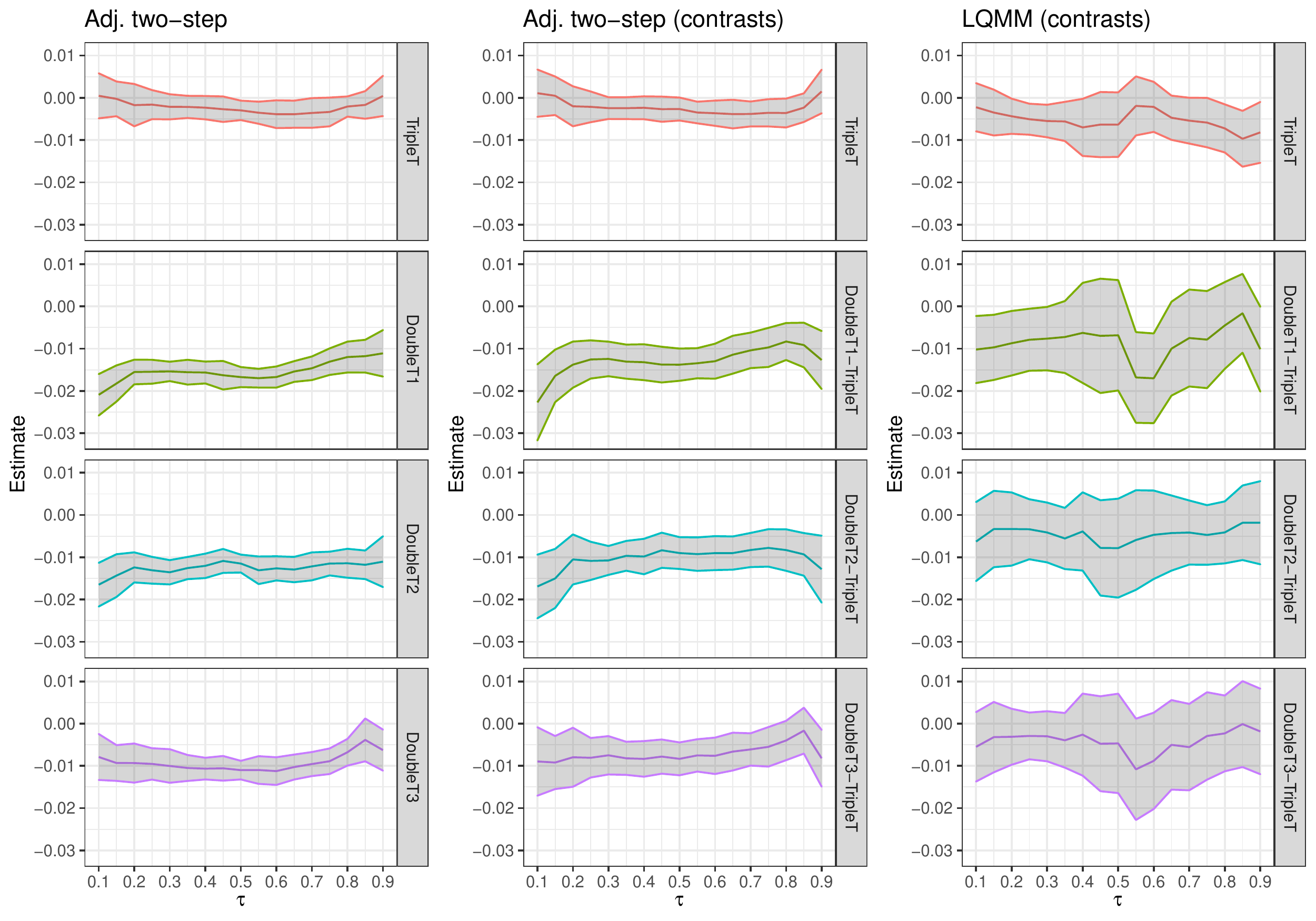}
           \caption{Estimated coefficients and 95\% confidence bands at varying $\tau$ for model \eqref{eq:model_separate_treats}. The left panels show results for slope coefficients $\beta^\tau_{1,h}$ ($h=1,\ldots,4$, adjusted two-step method) whereas the central and right panels show results for contrasts with triple therapy as reference (adjusted two-step method in the centre, LQMM to the right).}
    \label{fig:estimated_treatments_effects}
    \end{center}
	\end{figure}

While these results are interesting, we acknowledge one aspect of the data that our analysis does not account for: missing data. Out of the 1187 patients in the study, only 795 of them have measurements past the 30th week since their baseline. 
Missing data is not uncommon in ACTG studies and previous quantile regression analyses with longitudinal data have approached the problem by incorporating weights into the estimating equations \citep{lipsitz1997quantile}, employing hierarchical Bayesian models \citep{doi:10.1002/sim.7092,10.1093/biomet/asr052}, or by considering a linear quantile mixed hidden Markov model with a missing data indicator \citep{doi:10.1177/0962280216678433}. Incorporation of such methods falls beyond the scope of this paper, but could be an interesting avenue for future research. 

  \color{black}
\section{Discussion}\label{sec:conclusion}
\noindent
We have identified a gap in the literature concerning mixed effects models for quantile regression for clustered data:\ existing estimation methods may yield severely biased estimators for fixed effects parameters in situations with many, but small clusters. In this paper, we
 propose a new estimation method that relies on predicted random effects computed by using an LQMM working framework (in particular, at the quantile level of interest), standard quantile regression with offsets, and a bias-adjustment by means of a novel bootstrap sampling technique.
In the simulation study, the proposed estimator shows considerably smaller bias compared to the available competitors, especially in situations with small clusters. The RW adjustment appears to be particularly beneficial for estimating slope parameters, while the results are less clear for the intercept and could be studied further.
The two-step estimation procedure may be seen as the onset in an iterative procedure alternating between
estimation of the regression parameters for fixed random effects and prediction of random effects for fixed regression parameters. An ALD working model with random effects only (no fixed effects) can be used in the second step, and this requires minor modifications of the current implementation of  the \texttt{lqmm()} function.

Hitherto, the literature for quantile regression for clustered data has focused on studying asymptotics for increasing both the number of clusters and the cluster size \citep{koenker2004quantile, RePEc:eee:econom:v:170:y:2012:i:1:p:76-91, doi:10.1111/j.1368-423X.2011.00349.x, RePEc:cfr:cefirw:w0249}. In such case, the cluster-specific parameters are asymptotically ``eliminated'' as stated by \cite{doi:10.1111/j.1368-423X.2011.00349.x} or ``concentrated out'' as stated by \citet{RePEc:eee:econom:v:170:y:2012:i:1:p:76-91} and act as known quantities for the asymptotics of $\beta^\tau$.
On the other hand, the theoretical study of the estimators is inherently
challenging, when cluster size is fixed, and only the number of clusters increases to
  infinity. Results from (generalized) linear mixed models do not
  carry over for primarily two reasons. First, 
the criterion functions constructed from the check
  function is not differentiable. Second, the distributional assumptions are typically held to the minimum and focus on the relationship between the covariates and the quantile of interest. 
  In particular,
\cite{geraci2007quantile, geraci2014linear}
do not mention any attempts
  to derive asymptotic results for the LQMM estimator and rely on bootstrap methods for inference. 
Neither do we provide asymptotic results for our estimators, nor claim
that bias is \emph{removed} asymptotically. The main
difficulty lies in the prediction accuracy of the random effect predictors, which are used
as one of the main ingredients in the bootstrap sampling procedure. If
the predicted random effects do not accurately capture the variation of the cluster-specific random effects, then
the estimated bias may not represent the bias of the unadjusted
estimator. Therefore, when we are neither assuming an increasing cluster
size nor considering a specific data generating model, then it is
difficult to prove asymptotic results for our estimators, and we
leave this for future research.

\color{black}
  

Mean regression models for longitudinal data often incorporate more complex within-subject dependence structures than the one modeled by random intercepts alone (compound symmetry). Similar attempts do not seem to exist for quantile regression. The two-step estimator is not readily modified to take a serial dependence into account, but the RW bootstrap sampling could be easily adapted such as by sampling the weights for wild bootstrap at the subject level rather than at the measurement level. 
Moreover, longitudinal studies may involve drop-outs and occasional missing data, with data not missing at random, and how to incorporate such missingness in quantile regression in an appropriate way remains an open research problem.

One direction that the proposed methodology opens up is to  consider quantile
regression for time series data (one long series rather than many shorter
series), see \citet{Xiao2017}. In such case, the quantile model would be
$Q_{Y_t|X_t}(\tau) = X_t^T\beta^\tau + u_t^\tau$ where $Y_t$
and $X_t$ denote the response and covariate, respectively, at time
$t$ ($t=1,\ldots,T$), and $\{u_t^\tau\}_{t=1,\ldots,T}$  is a latent
series which describes (random) fluctuations of quantiles over time.
Another direction is to extend the approach to 
multi-level data with multiple levels of nested random effects or data with several, but non-nested random effects. {The ideas behind the methods from this paper (existing as well as our proposed method) would carry over to such situations, but a rigorous investigation of this extension is left for future research.}
\color{black}



\bibliographystyle{apalike}

\section*{Acknowledgements}

\noindent
The project was partly funded by the Danish Research Council  (DFF grant
7014-00221).We would like to thank the associate editor and two anonymous reviewers for suggestions and comments that improved the paper.

\bibliography{bibliography.bib,helle-bib}

\appendix
\section{Appendix} 
\noindent
The appendix contains additional numerical results from the simulation
study with data generated from model \eqref{eq:simmodel}. The results are  discussed  in the main text.
Tables~\ref{table:combi50_othermethods} and
\ref{table:combi10_othermethods} compare various existing approaches when
both the number of clusters and the
cluster size vary; other simulation parameters are specified by their level at the benchmark scenario.
Estimation is carried out for quantile levels
$\tau=0.5$ (Table \ref{table:combi50_othermethods}) and 
$\tau=0.1$ (Table \ref{table:combi10_othermethods}), respectively,
with results based on 200 replications. 
It is not possible to compute the jackknife estimator
when $n_i=3$ because clusters cannot be split into two subsets with
several observations per cluster. Furthermore, in the scenario with
$N=1000$, $n_i=12$ and $\tau=0.1$ there were convergence problems for the $\ell_1$-penalized estimator for
two datasets, and the results for this estimator are
based on the remaining 198 replications. 
Table~\ref{table:combi50} and Table~\ref{table:combi10} have the same
structure as described above and consider the same scenarios; they evaluate the performance of the
 LQMM estimator and our two proposed methods in 1000
 replications.
Notice the difference in the number of replications; as mentioned in
Section \ref{sec:compare} it is due to the computational burden of some of
the traditional estimators. 
\color{black}
Finally, Table~\ref{table:t3+ald} summarizes the results for the case when the error terms in \eqref{eq:simmodel} are either sampled from a scaled
$t$-distribution in the benchmark scenario, from an ALD distribution
in the benchmark scenario or an ALD distribution when $\gamma=0$. Results correspond to the quantile level $\tau=0.1$ and are based 
on 200 replications. 
\color{black}

\begin{table}[htb]
	 \begin{adjustbox}{angle=90}
	\centering
	\begin{tabular}{cclcccccccccccc}
		& & &\multicolumn{6}{c}{$\beta_0^\tau$} & \multicolumn{6}{c}{$\beta_1^\tau$} \\
          $N$ & $n_i$ & & oracle &Canay's & jackknife & $\ell_1$-pen & $\ell_2$-pen & marg & oracle &Canay's & jackknife & $\ell_1$-pen & $\ell_2$-pen & marg\\ \hline

		& &Bias  & $<0.01$ & 0.01 & --- & 0.01 & 0.01 & 0.01 & -0.01 &-0.02 & --- &  -0.02 & -0.02 & -0.02\\
		$500$& 3 & SD & 0.07 &0.08 &--- &0.09& 0.09& 0.09& 0.14&0.15 & ---& 0.15 &0.15 & 0.17\\
		& & RMSE & 0.07 &0.08 &---& 0.09& 0.09& 0.09 & 0.14&0.15 &---& 0.15 & 0.15 & 0.17\\
		\hline
		
                & &Bias & $<0.01$  & 0.01& ---& 0.01& 0.01& 0.01&$<0.01$  &$<0.01$ & --- & -0.01 & -0.01 & -0.01\\
		$1000$& 3 & SD &0.05& 0.06 & ---& 0.07& 0.07& 0.07 & 0.09& 0.10 & --- & 0.11 & 0.11 &0.12\\
		& & RMSE & 0.05 &0.06 &--- &0.07& 0.07& 0.07& 0.09&0.10 &--- & 0.11 & 0.11 & 0.12\\
		\hline

                & &Bias & $<0.01$ & $<0.01$& -0.01& $<0.01$& $<0.01$& 0.01 & -0.01&0.01 &0.01 & -0.01 & $<0.01$ &  -0.02\\
		$500$& 6 & SD & 0.05&0.07 &0.11& 0.07& 0.07& 0.08& 0.11 &0.10 & 0.13 &  0.11 &  0.11&  0.13\\
		& & RMSE & 0.05&0.07&  0.11& 0.07& 0.07& 0.08& 0.11&0.10 & 0.13 & 0.11 & 0.11 & 0.13\\
		\hline
		
		& & Bias &$<0.01$ & $<0.01$& $<0.01$& $<0.01$& $<0.01$& $<0.01$ & -0.01 &$<0.01$ & -0.01 & $<0.01$ & -0.01 & -0.01\\
		$1000$& 6&  SD & 0.03 & 0.05 & 0.08& 0.05& 0.05& 0.05& 0.07&0.07 & 0.08 &  0.07 & 0.08 & 0.09\\
		& & RMSE & 0.03&0.05& 0.08& 0.05& 0.05& 0.05& 0.07&0.07 & 0.08 & 0.07 & 0.08 & 0.09\\
          \hline
          
		& &Bias & $<0.01$ & $<0.01$& $<0.01$& -0.01& $<0.01$& $<0.01$ & $<0.01$ &$<0.01$ & $<0.01$ &$<0.01$ &$<0.01$ &$<0.01$ \\
		$500$& 12 & SD&0.03 &0.06 &0.13& 0.07& 0.06& 0.07& 0.07& 0.07 & 0.09 & 0.07 & 0.07 & 0.09\\
		& & RMSE & 0.03& 0.06 &0.13& 0.07& 0.06& 0.07& 0.07&0.07 & 0.09 & 0.07 & 0.07 & 0.09\\
		\hline
		
                & &Bias &$<0.01$ &$<0.01$& $<0.01$& $<0.01$& $<0.01$& 0.01 &$<0.01$&$<0.01$ & $<0.01$ & $<0.01$ & $<0.01$ & $<0.01$\\
		$1000$& 12 & SD &0.03 & 0.04& 0.10& 0.05& 0.04& 0.05& 0.05
&0.05 & 0.06 & 0.05 & 0.05 &0.06\\
		& & RMSE & 0.03 &0.04 &0.10& 0.05& 0.04& 0.05& 0.05&0.05 & 0.06 & 0.05 & 0.05 & 0.06\\
		\hline

	\end{tabular}
	
	\end{adjustbox}
	\caption{Bias, standard deviation, and RMSE for the oracle,
          Canay's, the jackknife, the $\ell_1$-penalized, the
          $\ell_2$-penalized and the marginal estimators. The quantile
          level is $\tau=0.5$, and results are based on 200 replications.}
	\label{table:combi50_othermethods}
	\end{table}

\begin{table}[htb]
	 \begin{adjustbox}{angle=90}
	\centering
	\begin{tabular}{cclcccccccccccc}
		& & &\multicolumn{6}{c}{$\beta_0^\tau$} & \multicolumn{6}{c}{$\beta_1^\tau$} \\
          $N$ & $n_i$ & & oracle &Canay's & jackknife & $\ell_1$-pen & $\ell_2$-pen & marg & oracle &Canay's & jackknife & $\ell_1$-pen & $\ell_2$-pen & marg\\ \hline

		& &Bias & $<0.01$ & 0.16 &  ---  & -0.42 & -0.48 & -0.51 &$<0.01$ &0.23 &  ---  &  0.10 & 0.10 & 0.09\\
		$500$& 3 & SD & 0.10 & 0.10 &---  &0.14& 0.12& 0.13& 0.18 & 0.17 & --- & 0.21 &0.20 & 0.23\\
		& & RMSE & 0.10 & 0.19 &--- &0.44& 0.50 & 0.53 & 0.18 &0.29 &--- & 0.23 & 0.23 & 0.25\\
		\hline
		
                & &Bias  & $<0.01$ & 0.16& --- & -0.38& -0.46& -0.52 & $<0.01$ & 0.24 & ---  & 0.11 & 0.12 & 0.11\\
		$1000$& 3 & SD & 0.07 & 0.07 & --- & 0.10 & 0.09& 0.10 &0.14 & 0.13 & --- & 0.15 & 0.16 &0.18\\
		& & RMSE & 0.07 & 0.18 &--- &0.40& 0.47& 0.53& 0.13 & 0.27 &---  & 0.19 & 0.20 & 0.21\\
		\hline

                & &Bias & 0.01 & 0.07& -0.15& -0.10& -0.36 & -0.53 & -0.01 & 0.11 &0.05 & 0.09 & 0.10 & 0.12\\
		$500$& 6 & SD & 0.07 & 0.08 &0.22& 0.13& 0.09& 0.10& 0.13 &0.13 & 0.21&  0.14 &  0.14&  0.16\\
		& & RMSE & 0.07 & 0.11&  0.27& 0.17& 0.37& 0.54& 0.13 &0.17 & 0.22 & 0.17 & 0.17 & 0.20\\
		\hline
		
		& & Bias & $<0.01$  & 0.07& -0.14& -0.02& -0.34& -0.53 & $<0.01$  & 0.13 &0.03 & 0.09 &0.10 & 0.12\\
		$1000$& 6&  SD & 0.04 & 0.06 & 0.19& 0.10& 0.06& 0.07& 0.09 &0.10 & 0.16 &  0.10& 0.10 & 0.12\\
		& & RMSE & 0.04 &0.09& 0.23&  0.10 & 0.35& 0.53& 0.09 & 0.16 & 0.17 & 0.14 & 0.14 & 0.17\\
          \hline
          
		& &Bias  & $<0.01$&0.03& -0.06& -0.04& -0.34& -0.52 & $<0.01$ &0.06  & -0.01 & 0.07 & 0.10 &0.12 \\
		$500$& 12 & SD & 0.05 & 0.07 &0.21& 0.09& 0.08& 0.09& 0.10 & 0.09 & 0.15 & 0.09 & 0.10 & 0.12\\
		& & RMSE & 0.05 & 0.08&0.22& 0.10& 0.35& 0.53& 0.09 & 0.11& 0.15 & 0.12 & 0.14 & 0.17\\
		\hline
		
                & &Bias & $<0.01$ & 0.04& -0.04& -0.03& -0.34 & -0.52 & $<0.01$ &0.06 & $<0.01$ & 0.07 & 0.10 & 0.12\\
		$1000$& 12 & SD & 0.03 & 0.05& 0.16&  0.08& 0.06& 0.06&0.07 &0.07 & 0.10 & 0.07 & 0.07 &0.09\\
		& & RMSE & 0.03 & 0.06 &0.17&0.08& 0.34& 0.52& 0.06 & 0.09 & 0.10 & 0.10 & 0.12 & 0.15\\
		\hline

	\end{tabular}
	
	\end{adjustbox}
	\caption{Bias, standard deviation, and RMSE for the oracle,
          Canay's, the jackknife, the $\ell_1$-penalized, the
          $\ell_2$-penalized and the marginal estimators. The quantile
          level is $\tau=0.1$, and results are based on 200 replications.}
	\label{table:combi10_othermethods}
	\end{table}

\begin{table}[htb]
	\centering
	\begin{tabular}{cclcccccc}
		& & &\multicolumn{3}{c}{$\beta_0^\tau$} & \multicolumn{3}{c}{$\beta_1^\tau$} \\
          $N$ & $n_i$ & & lqmm & two-step & adj (RW) & lqmm & two-step & adj (RW)\\ \hline

		& &Bias  & $<0.01$& $<0.01$& $<0.01$& $<0.01$& $<0.01$ &$<0.01$\\
		$500$& 3 & SD & 0.09 &0.09 &0.09& 0.14& 0.15& 0.15\\
		& & RMSE & 0.09 &0.09& 0.09& 0.14& 0.15 &0.15\\
		\hline
		
                & &Bias  & $<0.01$& $<0.01$& $<0.01$& $<0.01$& $<0.01$ &$<0.01$\\
		$1000$& 3 & SD & 0.07& 0.06& 0.07& 0.10& 0.11& 0.11\\
		& & RMSE & 0.07 &0.06 &0.07& 0.10& 0.11& 0.11\\
		\hline

                & &Bias  & $<0.01$& $<0.01$& $<0.01$& $<0.01$& $<0.01$ &$<0.01$\\
		$500$& 6 & SD &0.07 &0.06& 0.07& 0.10& 0.10& 0.10\\
		& & RMSE &0.07& 0.06& 0.07& 0.10& 0.10& 0.10\\
		\hline
		
		& & Bias & $<0.01$& $<0.01$& $<0.01$& $<0.01$& $<0.01$ &$<0.01$\\
		$1000$& 6&  SD & 0.06 &0.05& 0.05& 0.07& 0.07& 0.07\\
		& & RMSE &0.06& 0.05& 0.05& 0.07& 0.07& 0.07\\
          \hline
          
		& &Bias  & $<0.01$& $<0.01$& $<0.01$& $<0.01$& $<0.01$ &$<0.01$\\
		$500$& 12 & SD & 0.09 &0.06& 0.06& 0.07& 0.07& 0.07\\
		& & RMSE & 0.09 &0.06& 0.06& 0.07& 0.07& 0.07\\
		\hline
		
                & &Bias & $<0.01$& $<0.01$& $<0.01$& $<0.01$& $<0.01$ &$<0.01$\\
		$1000$& 12 & SD & 0.06& 0.04& 0.04& 0.05& 0.05& 0.05\\
		& & RMSE & 0.06 &0.04& 0.04& 0.05& 0.05& 0.05\\
		\hline

	\end{tabular}
	\caption{Bias, standard deviation, and RMSE for the LQMM estimator (lqmm), the two-step estimator (two-step), and bootstrap-adjusted two-step estimator (adj) where bootstrap samples are generated with the RW method. The
quantile level is $\tau=0.5$,  and results are
               based on 1000 replications.}
	\label{table:combi50}
      \end{table}

\begin{table}[htb]
	\centering
	\begin{tabular}{cclcccccc}
		& & &\multicolumn{3}{c}{$\beta_0^\tau$} & \multicolumn{3}{c}{$\beta_1^\tau$} \\
          $N$ & $n_i$ & & lqmm & two-step & adj (RW) & lqmm & two-step & adj (RW)\\ \hline

		& &Bias & 0.02 &0.09& 0.06& 0.25& 0.10& 0.05\\
		$500$& 3 & SD & 0.16 &0.11& 0.14& 0.21& 0.20& 0.23\\
		& & RMSE & 0.16 &0.15& 0.15& 0.33& 0.22& 0.23\\
		\hline
		
                & &Bias &0.04 &0.09& 0.05& 0.26& 0.10& 0.05\\
		$1000$& 3 & SD &0.11 &0.08& 0.10& 0.15& 0.14 &0.16\\
		& & RMSE & 0.12& 0.12& 0.12& 0.30& 0.18& 0.17\\
		\hline

		& &Bias & -0.07& 0.03&  -0.02 &0.15& 0.06&  0.02\\
		$500$& 6 & SD & 0.13&0.08& 0.10& 0.15 &0.14& 0.15\\
		& & RMSE & 0.15&0.09& 0.10 &0.21&0.15& 0.16\\
		\hline
		
		& & Bias &-0.05 &0.03& -0.03&0.15&0.05& 0.02 \\
		$1000$& 6&  SD & 0.10&0.06 & 0.07 & 0.10&0.09 & 0.11 \\
		& & RMSE &0.11&0.07& 0.08& 0.18&0.11& 0.11\\
          \hline
          
		& &Bias & -0.06 & 0.01& -0.05&  0.08& 0.03& $<0.01$\\
		$500$& 12 & SD & 0.12& 0.07& 0.07& 0.10& 0.10& 0.11\\
		& & RMSE & 0.13& 0.07& 0.09& 0.12& 0.10& 0.11\\
		\hline
		
                & &Bias & -0.05  &0.01& -0.05& 0.07& 0.03& $<0.01$ \\
		$1000$& 12 & SD & 0.09& 0.05& 0.05& 0.07& 0.07 &0.08\\
		& & RMSE& 0.11& 0.05& 0.07& 0.10 &0.07& 0.08\\
		\hline

	\end{tabular}
	\caption{Bias, standard deviation, and RMSE for the LQMM
          estimator (lqmm), the two-step estimator (two-step), and
          bootstrap-adjusted two-step estimator (adj) where bootstrap
          samples are generated with the RW method. The quantile level
          is $\tau=0.1$, and results are
               based on 1000 replications. }
	\label{table:combi10}
\end{table}

\begin{table}[htb]
	\centering
	\begin{tabular}{ccclcccccc}
		&&& & \multicolumn{3}{c}{$\beta_0^\tau$} & \multicolumn{3}{c}{$\beta_1^\tau$} \\
		$e_{ij}$&$N$&$n_i$ & & lqmm & two-step & adj (RW) & lqmm & two-step & adj (RW)\\ \hline
		
		
		
	&&& Bias & -0.27&-0.04&-0.15&0.14 & 0.00 & 0.00 \\
		$t_3$&$500$& 6& SD &0.14&0.08&0.10&0.12& 0.11&0.14\\
		&&& RMSE & 0.30&0.09&0.18&0.18&0.12&0.14\\
		\hline

		&&& Bias & -0.25&-0.05&-0.16&0.15&0.06&0.01\\
		$t_3$&$1000$&6& SD &  0.11&0.06&0.07&0.09&0.09&0.12\\
		&&& RMSE &0.28&0.07&0.17&0.17&0.11&0.12\\
		\hline

          		&&& Bias & -0.18&-0.05&-0.15&0.10&0.04&0.02 \\
		$t_3$&$500$&12& SD& 0.13&0.07&0.08&0.08&0.09&0.11 \\
		&&& RMSE &0.22&0.08&0.17&0.13&0.10&0.11\\
		\hline
		\hline

		&&& Bias &  -0.12& -0.10 &-0.05&0.06 & 0.05 & 0.04\\
		 ALD&$500$& 6& SD &0.13&0.06&0.07&0.06& 0.08&0.09\\
		&&& RMSE & 0.17&0.12&0.09& 0.08& 0.09&0.10\\
		\hline
				
		&&& Bias &  -0.08&-0.10&-0.06&0.05&0.05&0.04\\
		ALD&$1000$&6& SD &  0.10&0.04&0.05&0.05&0.05&0.06\\
		&&& RMSE &0.12&0.11&0.07&0.07&0.07&0.07\\
          \hline

          		&&& Bias & -0.08& -0.07 & -0.03 &0.03&0.03&0.02 \\
		ALD &$500$&12& SD& 0.12&0.05&0.05&0.04&0.04&0.05 \\
		&&& RMSE &0.15&0.09&0.06&0.05&0.05&0.05\\
		\hline
		\hline
		&&& Bias & -0.13&-0.06&-0.04&0.00 & 0.00 & -0.01 \\
		ALD&$500$& 6& SD &0.14&0.05&0.06&0.05& 0.06&0.07\\
		$(\gamma=0)$ &&& RMSE & 0.19&0.08&0.07&0.05&0.06&0.07\\
		\hline

		&&& Bias & -0.07&-0.07&-0.04&0.00&0.00&0.00\\
		ALD&$1000$&6& SD &  0.11&0.04&0.05&0.04&0.04&0.05\\
		$(\gamma=0)$ &&& RMSE &0.13&0.08&0.06&0.04&0.04&0.05\\
		\hline

        &&& Bias & -0.07&-0.05&-0.02&0.00&0.00&0.00 \\
		ALD&$500$&12& SD& 0.13&0.05&0.05&0.03&0.04&0.04 \\
		$(\gamma=0)$ &&& RMSE &0.14&0.07&0.06&0.03&0.04&0.04\\
		\hline

	\end{tabular}
	\caption{Bias, standard deviation, and RMSE for the LQMM
		estimator (lqmm), the two-step estimator
		(two-step), and bootstrap-adjusted two-step estimator (adj)
		where bootstrap samples are generated with the RW method.
		The residuals 
		are sampled from a scaled $t_3$ when $\gamma=0.4$ (top part), and from an ALD when either $\gamma=0.4$ (central part) or $\gamma=0$ (bottom part).
		The quantile level is $\tau=0.1$, and results are
               based 200 replications.}
           \label{table:t3+ald}
\end{table}

\end{document}